\documentclass{acm_proc_article-sp}

\usepackage[utf8]{inputenc}
\usepackage[T1]{fontenc}
\usepackage{cite}
\usepackage{mathtools}
\usepackage{tikz}
\usepackage{xspace}
\usepackage{drawmatrix}
\usepackage{booktabs}
\usepackage{multirow}
\usepackage{listings}
\usepackage{siunitx}
\usepackage{hyperref}

\usepackage{aicescover}

\colorlet{graybg}{gray!10}
\colorlet{plot1}{red}
\colorlet{plot2}{green!75!black}
\colorlet{plot3}{blue}
\colorlet{plot4}{yellow!75!black}
\colorlet{plot5}{cyan!75!black}
\colorlet{plot6}{magenta!75!black}

\hypersetup{
    colorlinks=true,       
    linkcolor=blue!50!black,          
    citecolor=green!50!black,        
    filecolor={.5 0 0},      
    urlcolor=red!50!black,           
    hypertexnames=false,
}

\usetikzlibrary{
    external,
    scopes,
    positioning,
    calc
}

\lstdefinelanguage{experiment}{
    keywords=[1]{for, repeat, times, sum, over, in, parallel},
    keywordstyle=[1]{\bf},
}
\newcounter{experiment}
\newsavebox\experimentcodebox%
\newenvironment{experiment}[4]{%
    \newcommand\expname{#1}%
    \newcommand\sysname{#2}%
    \newcommand\blasname{#3}%
    \newcommand\data{#4}%
    \newcommand\drawskip{\vspace{-\baselineskip}\vspace{.5\medskipamount}}
    \stepcounter{experiment}
    \begin{lrbox}\experimentcodebox%
}{
    \end{lrbox}%
    {
        \scriptsize
        \setlength{\fboxsep}{0pt}%
        \colorbox{graybg}{%
            \begin{tabular}{|l|l|r|}
                \hline
                \tt \sysname & \tt \blasname 
                &Experiment \theexperiment: \expname\\
                \hline
                \multicolumn{3}{|l|}{
                    \usebox\experimentcodebox
                }\\
                \hline
                \multicolumn{3}{|p{\linewidth - 2\medskipamount}|}{
                    \vspace{-5pt}
                    \data
                }\\
                \hline
            \end{tabular}%
        }%
    }%
}

\newcommand\elaps{ELAPS\xspace}
\newcommand\sampler{{\sc Sampler}\xspace}
\newcommand\playmat{{\sc PlayMat}\xspace}
\newcommand\viewer{{\sc Viewer}\xspace}

\newcommand\parsum[1]{\textcolor{orange}{#1}\\}

\newcommand\pdj[1]{\textcolor{blue}{\bf Pdj: #1}}
\newcommand\ep[1]{\textcolor{green!75!black}{\bf EP: #1}}

 \renewcommand\parsum[1]{}
 \renewcommand\pdj[1]{}
 \renewcommand\ep[1]{}

\conferenceinfo{SC15}{'15 Austin, Texas USA}
\title{The \elaps Framework:\\ Experimental Linear Algebra Performance Studies}
\numberofauthors{2}
\author{
    \alignauthor Elmar Peise and Paolo Bientinesi\\
    \affaddr{AICES, RWTH Aachen}
    \email{\tt \{peise,pauldj\}@aices.rwth-aachen.de}
}
\aicescoverauthor{
    Elmar Peise and Paolo Bientinesi
}

\begin{document}
    \aicescoverpage
    \maketitle

    \begin{abstract}

    Optimal use of computing resources requires extensive coding, tuning and
    benchmarking. To boost developer productivity in these time consuming
    tasks, we introduce the Experimental Linear Algebra Performance Studies
    framework (\elaps), a multi-platform open source environment for fast yet
    powerful performance experimentation with dense linear algebra kernels,
    algorithms, and libraries. \elaps allows users to construct experiments to
    investigate how performance and efficiency vary depending on  factors such
    as caching, algorithmic parameters, problem size, and parallelism.
    Experiments are designed either through Python scripts or a specialized
    GUI, and run on the whole spectrum of architectures, ranging from laptops
    to clusters, accelerators, and supercomputers.  The resulting experiment
    reports provide various metrics and statistics that can be analyzed both
    numerically and visually.  We demonstrate the use of \elaps in four
    concrete application scenarios and in as many computing environments,
    illustrating its practical value in supporting critical performance
    decisions.
   
\end{abstract}

\terms{Performance, Experimentation}
\keywords{performance experiments, dense linear algebra}


    \section{Introduction}
    \label{sec:intro}
    \parsum{problem: time consuming performance optimization}
The field of high performance computing is largely concerned with the optimal
usage of available resources.  Since performance depends on the choice of
algorithms, parameters, libraries and even computing environment, maximizing
efficiency is a task that comes at the cost of extensive coding, tuning and
benchmarking.  To facilitate and support such time-consuming and repetitive
activities within the development of dense linear algebra software, we propose
a rich and flexible environment for rapid performance experimentation. 

\parsum{\elaps}
The Experimental Linear Algebra Performance Studies framework (\elaps) allows
users to create experiments for investigating how performance and efficiency
depend on factors such as caching, algorithmic parameters, problem size, and
parallelism.  Experiments are designed by combining one or more  algorithmic
constructs commonly encountered in linear algebra computations, and built
either through Python scripts or a specialized and intuitive GUI.  They then
can be executed either  locally or through batch-job systems, on hardware
ranging from laptops and accelerators to clusters and supercomputers.  Finally,
the results can be visualized and analyzed interactively, in terms of various
performance metrics and statistics.

\parsum{exmaples}
As demonstrated in this paper by means of examples raising in actual
applications, insights gained through \elaps serve as a solid ground to make
performance relevant design decisions.

\parsum{Structure of the paper}
The remainder of this paper is structured as follows: \autoref{sec:experiment}
introduces the experimental features supported by \elaps; the framework,
together with its structure and implementation are described in
\autoref{sec:elaps}.  Finally, \autoref{sec:examples} demonstrates the use of
\elaps as a decision-making aid in a series of application examples.

\subsubsection*{Related Work}
\parsum{tuning examples}
Performance optimization is a widespread activity, impacting virtually all
scientific computing disciplines; out of many works, here we mention three
examples in the field of linear algebra that are aligned with the studies
enabled by \elaps: the optimization of the algorithmic block size for LAPACK's
routines~\cite{lapacktuning}, the study of symmetric tridiagonal
eigensolvers~\cite{eigensolvers}, and the construction of algorithms for the
inversion of symmetric positive definite matrices~\cite{spdinv}.

\parsum{autotuning}
A popular approach for performance optimization is the ``auto-tuning'': On the
one hand, domain-specific libraries such as ATLAS~\cite{atlas} and {\sc
FFTW}~\cite{fftw} perform an automatic search (with or without explicit timing)
to deliver hardware-specific code; on the other hand, general-purpose languages
and libraries such as {\sc Active Harmony}~\cite{activeharmony}, {\sc
Atune-IL}~\cite{atuneil}, and {\sc Chapel}~\cite{chapel} make the exploration
of a parameter space an integral part of the computing environment.  A solution
that combines automation with machine learning techniques to offer on-line
selection of algorithms is proposed in \cite{machinelearning1}.  In constrast
with automated solutions, \elaps' objective is to enable  interactive and
insightful experimentation.

\parsum{tools}
Many application-level tools for profiling and analyzing existing codes exist
(examples include {\sc PAPI}~\cite{papi}, {\sc Tau}~\cite{tau}, {\sc
Vampir}~\cite{vampir}, and {\sc Scalasca}~\cite{scalasca}); collectively, they
offer support for the whole range of architectures, from single computing nodes
to large distributed computers.  In its curret form, \elaps targets
shared-memory platforms, and early experimentation.


    \section{Experiments}
    \label{sec:experiment}
    \parsum{concept of experiment}
While performance experiments come in all kinds of shapes and sizes, many of
them can be described by a few common features.  Within the \elaps framework,
we combine and generalize such features to provide a versatile central concept
of ``experiment''.  In this section, we discuss these basic features guided by
deliberately simple examples.  More complicated examples arising in actual
applications are then presented in \autoref{sec:examples}.

\parsum{most basic experiment: 1 Gemm}
We begin with a most elementary experiment: Measuring the performance of the
matrix-matrix product kernel {\tt dgemm}. 
  
\begin{experiment}{{\tt dgemm}}{E5-2670}{OpenBLAS}{$
    A, B, C \in \mathbb R ^{\num{1000} \times \num{1000}}
$}
    \begin{lstlisting}
#threads = 1
dgemm: !$
    \drawmatrix C \coloneqq
    \drawmatrix A \; 
    \drawmatrix B
$!
!\drawskip!
    \end{lstlisting}
\end{experiment}

As shown schematically above, this experiment runs on one core of an Intel {\sc
SandyBridge E5-2670} processor, using the {\sc OpenBLAS}
library~\cite{openblas}, and executes the double precision kernel {\tt dgemm}
once on random square matrices of size \num{1000}. Although simple, similar
experiments are commonly used to determine the attainable peak performance of a
given processor.

\parsum{metrics}
When combined with additional information on the hardware and the kernel's
complexity, the raw timing (in cycles) from this experiment leads to a number
of metrics, which yield more insights into how efficiently the CPU is used.
\begin{center}
    \begin{tabular}{ll}
        \toprule
        metric                  &value\\
        \midrule
        \si{cycles}             &\num{272551028} \\
        time [\si{ms}]          &\num{104.8} \\
        \si{Gflops/s}           &\num{19.1} \\
        \si{flops/cycle}        &\num{7.3} \\
        efficiency [\si{\%}]    &\num{91.7} \\
        \bottomrule
    \end{tabular}
\end{center}
Furthermore, if available, the {\sc Performance Application Programming
Interface} (PAPI)~\cite{papi} allows one to access useful hardware counters.
\begin{center}
    \begin{tabular}{lll}
        \toprule
        metric                      &counter name                       &value \\
        \midrule
        Level 1 cache misses        &{\tt PAPI\_L1\_TCM}                &\num{32933961} \\
        Conditional branch          &\multirow{2}{*}{\tt PAPI\_BR\_MSP} &\multirow{2}{*}{\num{3941}} \\
        instructions mispredicted \\
        \bottomrule
    \end{tabular}
\end{center}
With \elaps, all these metrics are readily available and easily extensible.

\subsection{Repetitions and Statistics}
Multiple executions of a kernel often result in fluctuating timings; the
reasons for such differences include library initialization overhead, cache
locality, and system jitter.  As customarily done, in \elaps this issue is
addressed by repeating each experiment several times, and by collecting
statistics.  As an example, let us consider an experiment that repeats the
kernel execution from Experiment 1 ten times on the same input matrices (i.e.,
the same memory locations):

\begin{experiment}{Repetitions}{E5-2670}{OpenBLAS}{$
    A, B, C \in \mathbb R ^{\num{100} \times \num{100}}
$}
    \begin{lstlisting}
#threads = 1
repeat 10 times:
    dgemm: !$
        \drawmatrix C \coloneqq
        \drawmatrix A \; 
        \drawmatrix B
    $!
!\drawskip!
    \end{lstlisting}
\end{experiment}

\begin{figure}[t]
    \includegraphics[width=\linewidth]{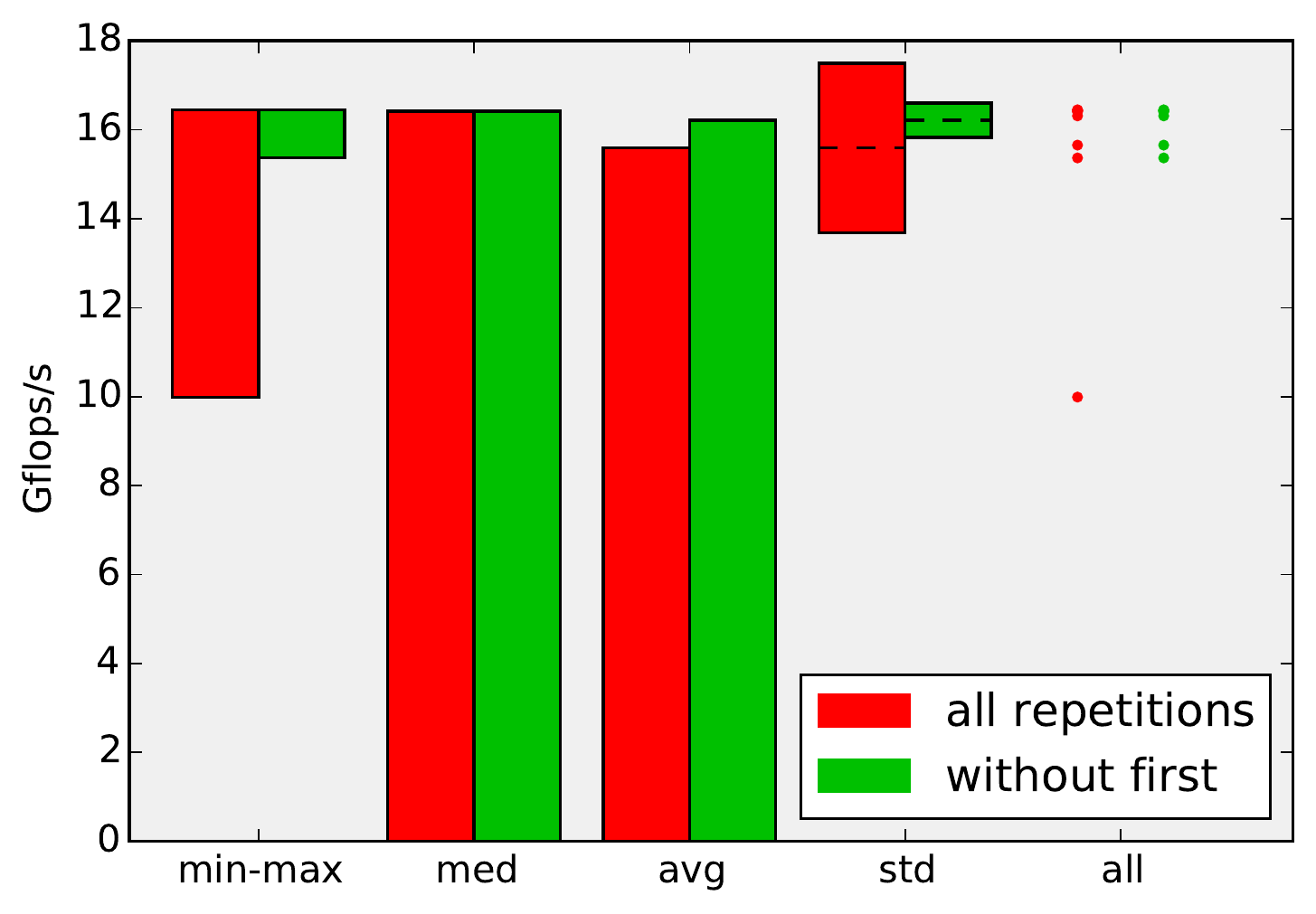}

    \caption{Performance statistics from 10 repetitions of {\tt dgemm}}
    \label{fig:ex:reps}
\end{figure}


\parsum{statistics}
This second experiment produces ten measurements, from which we derive
statistics as those presented in \autoref{fig:ex:reps}.  It is worth pointing
out that whenever multiple repetitions are executed and timed, the first one
almost inevitably represents an outlier; for the most part, this phenomenon is
connected to the initialization of the kernel library, but it is also due to
the loading and caching of data and instructions.  In general, a more accurate
representation of the effective performance is obtained by dropping the first
measurement of the lot.  In \autoref{fig:ex:reps} one can appreciate how
significantly the first repetition affects the various statistics, and most
noticeably, the minimum, the average and the standard deviation (std).

\parsum{always discard first from now on}
In order to avoid the impact of ``first-execution'' outliers, in all following
examples and studies we always discard the measurement relative to the first
repetition.

\subsection{Data Placement: Varying Operands}

\parsum{why vary matrices?}
In Experiment~2, the matrices $A$, $B$, and $C$ were reused across repetitions,
causing them to stay in cache; this scenario is also known as a ``warm data''.
Depending on the application, the assumption of warm data may or may not be
realistic; to reflect this in our experiments, we allow to ``vary'' the
operands (i.e., use different memory locations) individually in each
repetition.  Furthermore, in \elaps one can freely control the relative
position of varying operands: They can be stacked horizontally or vertically,
with or without an arbitrary offset. 

In the following experiment, while $A$ and $B$ are fixed and quite small, $C$
varies in each repetition (hence the subscript in $C_{\rm rep}$) and is
therefore never cached (``cold data'').

\begin{experiment}{Varying Operands}{E5-2670}{OpenBLAS}{$
    A \in \mathbb R^{\num{2000} \times \num{20}},
    B \in \mathbb R^{\num{20} \times \num{2000}},
    C_{\rm rep} \in \mathbb R^{\num{2000} \times \num{2000}}
$}
    \begin{lstlisting}
#threads = 1
repeat 100 times:
    dgemm: !$
        \drawmatrix{C_{\rm rep}} \coloneqq 
        \drawmatrix[width=.01] A \; 
        \drawmatrix[height=.01] B
    $!
!\drawskip!
    \end{lstlisting}
\end{experiment}

\begin{figure}[t]
    \includegraphics[width=\linewidth]{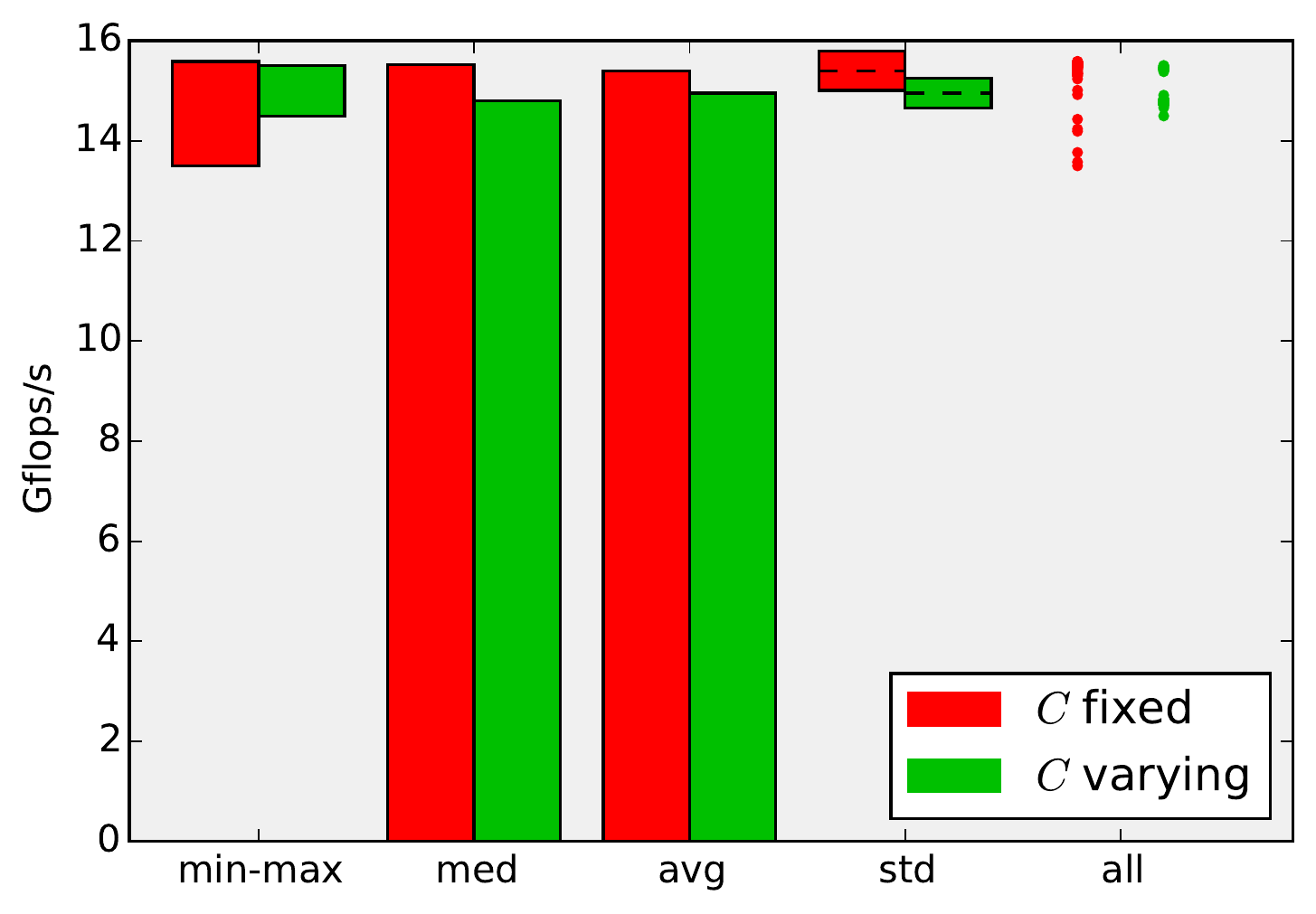}

    \caption{Influence of data locality on {\tt dgemm}}
    \label{fig:ex:vary}
\end{figure}

In \autoref{fig:ex:vary} we present the results of Experiment~3 and another
experiment in which the matrices $A$, $B$, and $C$ are all fixed.  The
performance loss due to the enforced out-of-cache scenario for $C$ is clearly
visible.

\subsection{Sequences of Kernels}
In addition to isolated kernels, \elaps allows to experiment with sequences of
calls.  Let us use the solution of a linear system as an example: The problem  
\begin{equation}
    \drawmatrix[width=.2]B \coloneqq \drawmatrix A^{-1} \drawmatrix[width=.2]B,
    \label{eqn:linsys}
\end{equation}
is typically solved by first LU-decomposing $A$ ({\tt dgetrf}), and then by
solving two triangular linear systems ({\tt dtrsm}).  The process---which is
also implemented in LAPACK's {\tt dgesv}---is replicated in
Experiment~4.\footnote{%
    For simplicity, we don't expose the pivoting vector and omit the row
    interchanging kernel {\tt dlaswp} that only contributes a lower order term
    to the execution time.
}

\begin{experiment}{Linear System Breakdown}{E5-2670}{OpenBLAS}{$
    A \in \mathbb R^{\num{1000} \times \num{1000}}, 
    B \in \mathbb R^{\num{1000} \times \num{200}}
$}
    \begin{lstlisting}
#threads = 1
repeat 10 times:
    dgetrf: !$
        \drawmatrix A \coloneqq {\rm LU}\left(\drawmatrix A\right)
    $!
    dtrsm: !$
        \drawmatrix[width=.2]B \coloneqq \drawmatrix[lower]A^{-1}
        \drawmatrix[width=.2]B
    $!
    dtrsm: !$
        \drawmatrix[width=.2]B \coloneqq \drawmatrix[upper]A^{-1}
        \drawmatrix[width=.2]B
    $!
!\drawskip!
    \end{lstlisting}
\end{experiment}

\begin{figure}[t]
    \includegraphics[width=\linewidth]{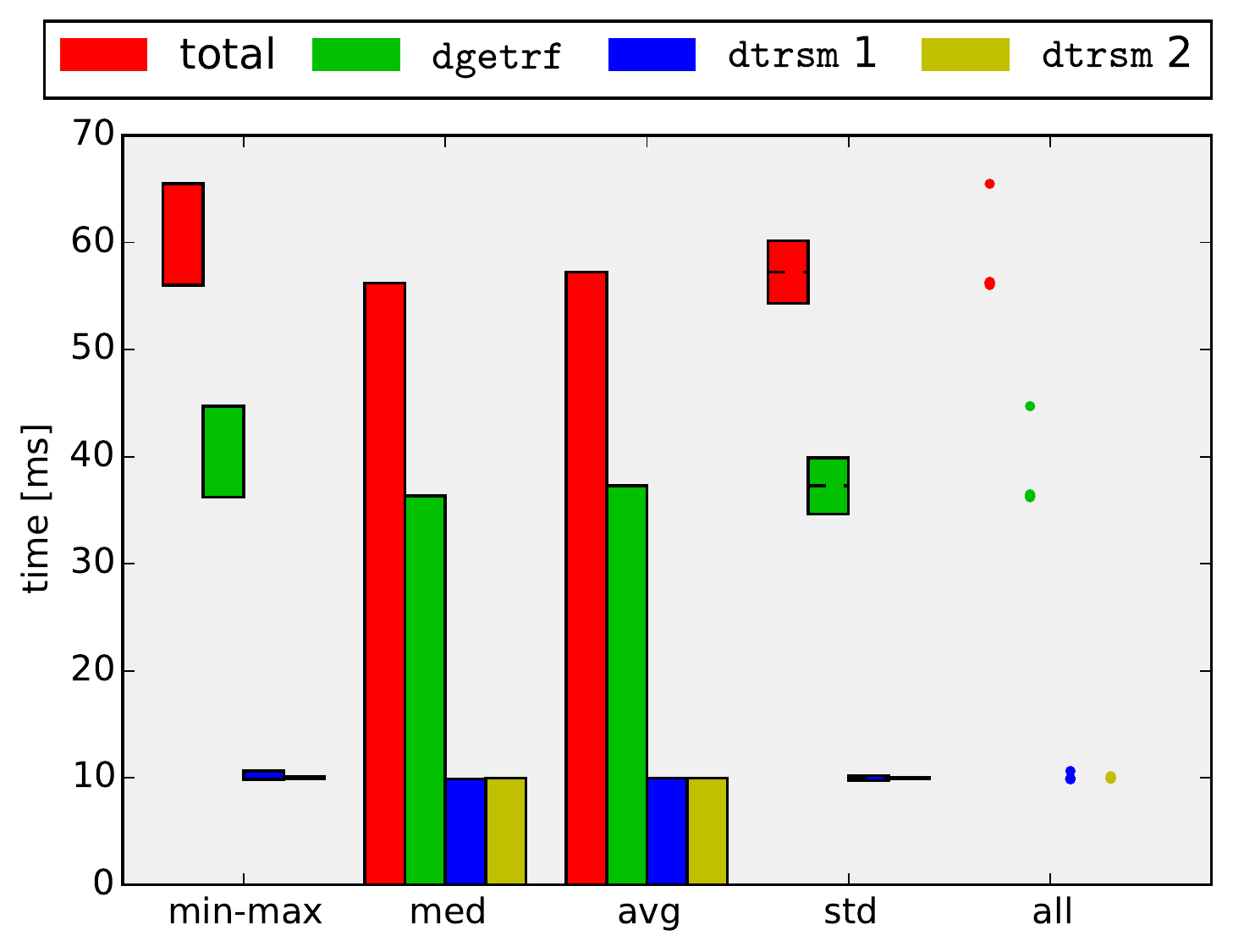}

    \caption{Breakdown of the timings for the solution of a linear system}
    \label{fig:ex:calls}
\end{figure}

\parsum{results}
For this experiment, \autoref{fig:ex:calls} shows both the total execution time
and the time spent in each individual kernel.  It is easy to realize that for
200 right-hand sides, the LU decomposition {\tt dgetrf} is responsible for more
than \SI{60}{\%} of the total execution time, while each of the {\tt dtrsm}'s
only contribute less than \SI{20}{\%}.

\subsection{Parameter Range}
\parsum{motivation and example}
So far we only considered experiments in which the sizes of the kernel operands
where fixed.  In many practical experiments however one wants to study the
performance of a routine over a range of parameters.  In the following example,
we use the routine {\tt dgesv}, which solves a linear system directly, to solve
problems of size $n$ with \num{500} right hand sides, where $n$ ranges from
\num{50} to \num{2000} in steps of \num{50}.

\begin{experiment}{Range}{E5-2670}{OpenBLAS}{$
    A \in \mathbb R^{n \times n}, 
    B \in \mathbb R^{n \times \num{500}}, 
$}
    \drawmatrixset{bbox style={fill=white, draw=gray, opacity=.5}}%
    \begin{lstlisting}
#threads = 1
for $n$ = 50:50:2000
    repeat 10 times:
        dgesv: !$
            \drawmatrix[bbox height=1, height=.025, width=.25]B \coloneqq
            \drawmatrix[bbox size=1, size=.025]A^{-1} \; 
            \drawmatrix[bbox height=1, height=.025, width=.25]B
        $!
!\drawskip!
    \end{lstlisting}
\end{experiment}

\begin{figure}[t]
    \includegraphics[width=\linewidth]{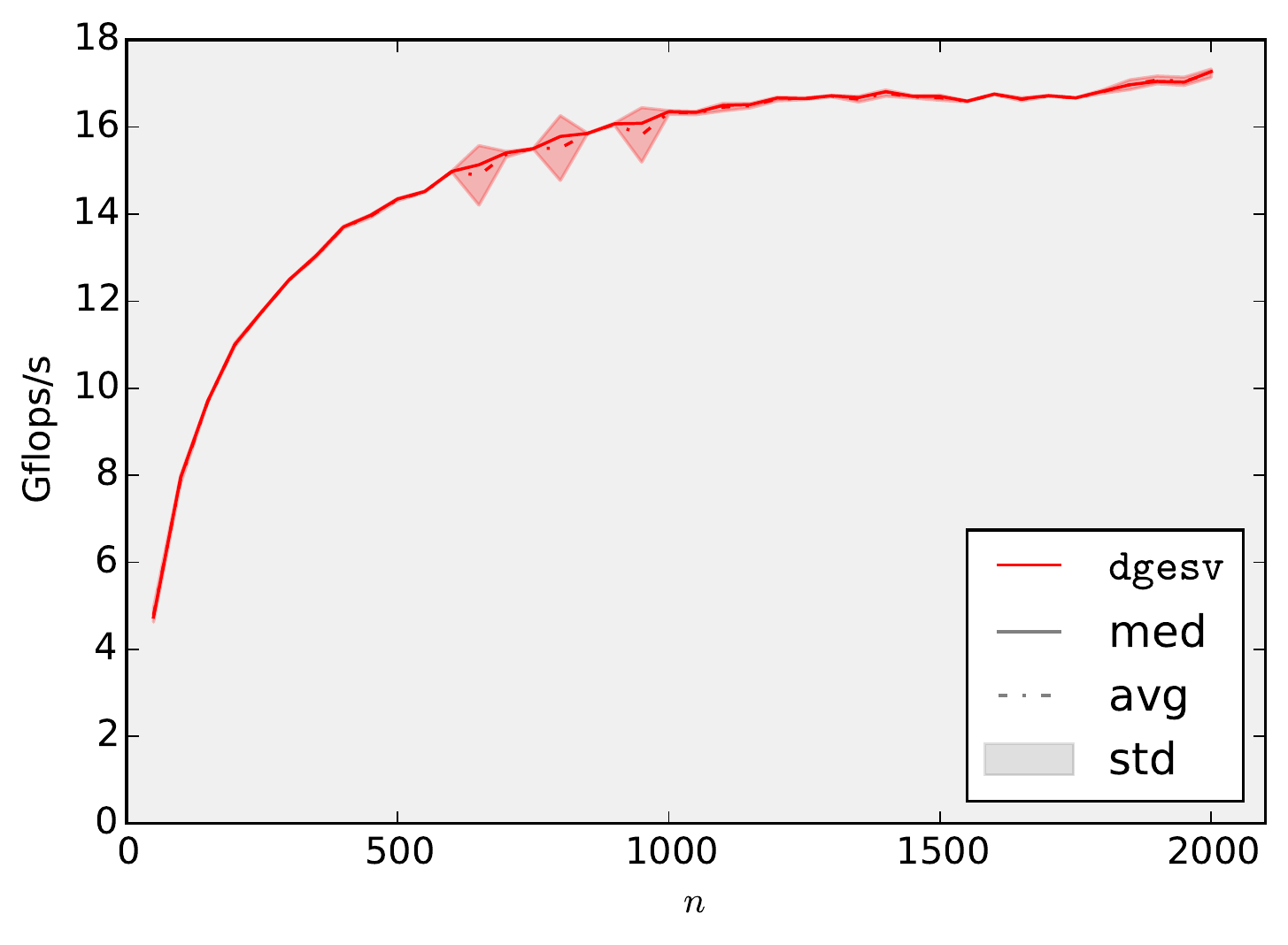}

    \caption{Solution of linear systems: performance.}
    \label{fig:ex:range}
\end{figure}


Performance results from Experiment~5 are shown in \autoref{fig:ex:range}; the
plot displays the increase in performance for increasing problem size, as
typical for dense linear algebra kernels.

\subsubsection{Threads Range}
\label{sec:ntrange}
Scalability studies are extremely common examples of experiments that make use
of ranges. In the following experiments, we compute the eigenvalue
decomposition of a symmetric matrix (of fixed size) using from \num{1} up to
\num{8} threads, and compare LAPACK's solvers {\tt dsyev}, {\tt dsyevx}, {\tt
dsyevr}, and {\tt dsyevd} (see \cite{lapack} for details on these routines).

\begin{experiment}{Threads Range}{E5-2670}{OpenBLAS}{$
    A, B \in \mathbb R^{\num{2000} \times \num{2000}}
$}
    \begin{lstlisting}
for $t$ = 1:8:
    #threads = $t$
    repeat 10 times:
        dsyev!$\ast$!: !$
            \drawmatrix A, \drawmatrix[width=0]\Lambda \coloneqq
            {\rm eig}\left(\drawmatrix A\right)
        $!
!\drawskip!
    \end{lstlisting}
\end{experiment}

\begin{figure}[t]
    \includegraphics[width=\linewidth]{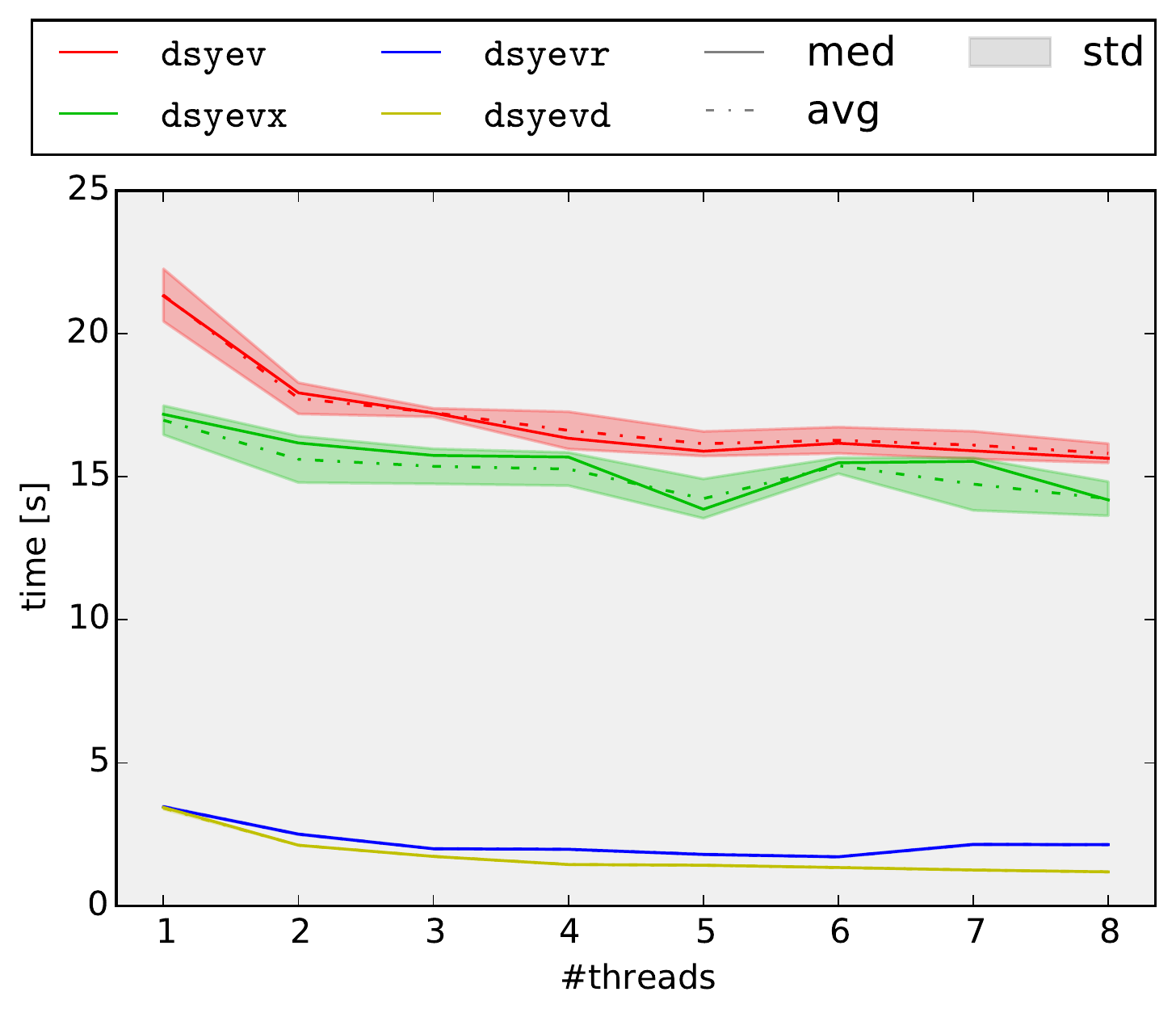}

    \caption{Scalability of LAPACK's symmetric dense eigensolvers on random matrices} 
    \label{fig:ex:ntrange}
\end{figure}


As one can appreciate from \autoref{fig:ex:ntrange}, \elaps makes it easy to
set up, execute, and compare the results of multiple experiments with varying
degrees of parallelism.

\subsection{Sum- and {\sc\large OpenMP}-Range}
\label{sec:sumrange}
\parsum{motivation and example}
In loop-based algorithms, the total execution time is often more meaningful
than an iteration-by-iteration break-down.  For this purpose, in addition to
the ``parameter range'' described in the previous subsection, \elaps also
provides a ``sum-range'', which yields the total contribution of the loop.  For
instance, the next experiment models the inversion of a lower triangular
matrix\footnote{%
    LAPACK's {\tt dtrtri} computes the inverse of a triangular matrix using a
    similar algorithm.
} of size \num{1000} via a blocked algorithm that traverses the matrix in steps
of a fixed block-size $nb$.

\begin{experiment}{Triangular Inversion}{E5-2670}{OpenBLAS}{$
    A00 \in \mathbb R^{j \times j},
    A10_{\downarrow j} \in \mathbb R^{\num{100} \times j},
    A11_j \in \mathbb R^{\num{100} \times \num{100}}
$}
    \begin{lstlisting}
#threads = 1
repeat 10 times:
    sum over $j$ = 0:$nb$:(1000-$nb$):
        dtrmm!\footnotemark!: !$
            {
                \drawmatrixset{bbox style={fill=white, draw=gray, opacity=.5}}
                \drawmatrix[bbox width=1, width=0, height=.111]{A10_{\downarrow j}^T}
                \coloneqq 
                \drawmatrix[bbox width=1, width=0, height=.111]{A10_{\downarrow j}^T} \;
            }
            \drawmatrix[lower, size=1, opacity=.5]{A00}
        $!
        dsyrk!\footnotemark!: !$
            \drawmatrix[lower, size=.111]{A11} \coloneqq
            \drawmatrix[lower, size=.111]{A11} -
            {
                \drawmatrixset{bbox style={fill=white, draw=gray, opacity=.5}}
                \drawmatrix[bbox width=1, width=0, height=.111]{A10_{\downarrow j}^T} \;
                \drawmatrix[bbox height=1, height=0, width=.111]{A10_{\downarrow j}^T}
            }
        $!
        dtrti2!\footnotemark!: !$
            \drawmatrix[lower, size=.111]{A11_j} \coloneqq
            \drawmatrix[lower, size=.111]{A11_j}^{-1}
        $!
!\drawskip!
    \end{lstlisting}
\end{experiment} 

\addtocounter{footnote}{-3}
\stepcounter{footnote}\footnotetext{%
    {\tt dtrmm}: Triangular matrix matrix multiplication.
}
\stepcounter{footnote}\footnotetext{%
    {\tt dsyrk}: Symmetric rank $k$ update.
}
\stepcounter{footnote}\footnotetext{%
    {\tt dtrti2}: Unblocked triangular inversion.
}

\begin{figure}[t]
    \includegraphics[width=\linewidth]{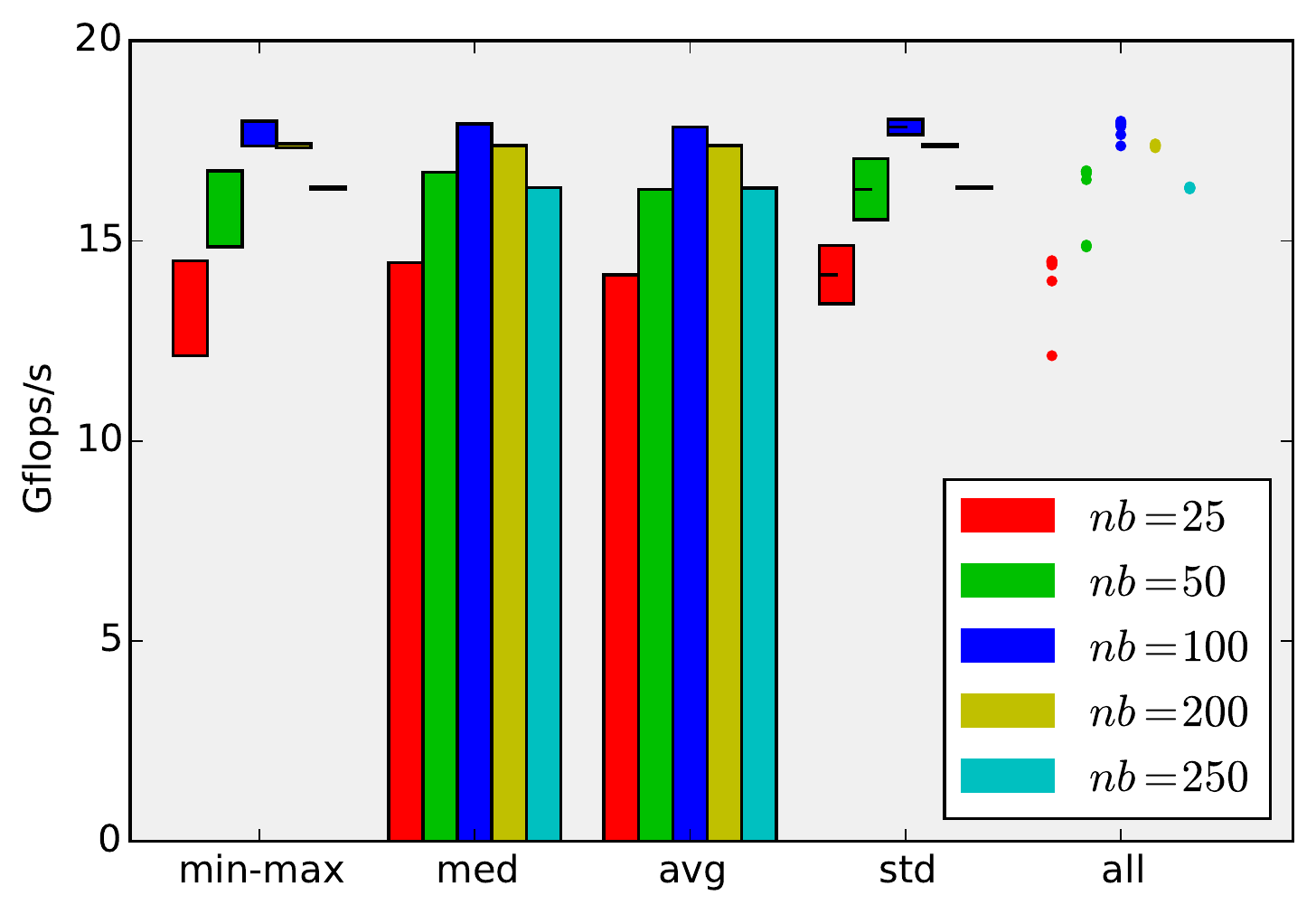}

    \caption{Influence of block-size on triangular inversion}
    \label{fig:ex:sumrange}
\end{figure}

\subsubsection{{\sc\large OpenMP}-Range}
\label{sec:ex-omprange}
\autoref{fig:ex:sumrange} reports the performance attained by this algorithm
for different block-sizes $nb$; the maximum is observed for $nb = \num{100}$.
The choice of parameters represents an important step to tailor algorithms for
a given architecture; for instance, the tuning of the block size is common to
many of the algorithms included in LAPACK~\cite{lapacktuning, lapack}. Notice
that a simpler and finer-grained experiment to optimize the block-size is
obtained by combining the sum-range with a parameter-range for $nb$.

\parsum{parallel sum range: same in parallel}
Multi-threading can typically be exploited in two different ways, namely,
either invoking a multi-threaded library (such as {\sc OpenBLAS}), or through
{\sc OpenMP}.  To investigate these alternatives, in the Experiments 8 and 9 we
implement the solution of a triangular linear system with a tall and skinny
right-hand side 1) with {\sc OpenBLAS}'s threaded {\tt dtrsm} kernel, and 2) as
a series of parallel {\tt dtrsv}'s\footnote{%
    {\tt dtrsv}: Linear system solve with a single right-hand side.
} using \elaps's {\sc OpenMP}-range.

\begin{experiment}{Tall and skinny {\tt dtrsm}}{E5-2670}{OpenBLAS}{$
    A \in \mathbb R^{\num{2000} \times \num{2000}},
    B \in \mathbb R^{\num{2000} \times \num{8}}
$}
    \begin{lstlisting}
#threads = 8
repeat 10 times:
    dtrsm: !$
        \drawmatrix[width=.004]B \coloneqq
        \drawmatrix[lower]A^{-1}
        \drawmatrix[width=.004]B
    $!
!\drawskip!
    \end{lstlisting}
\end{experiment}

\begin{experiment}{Parallel {\tt dtrsv}'s}{E5-2670}{OpenBLAS}{$
    A \in \mathbb R^{\num{2000} \times \num{2000}},
    b_i \in \mathbb R^{\num{2000}}
$}
    \begin{lstlisting}
#threads = 1
repeat 10 times:
    in parallel i = 1:8:
        dtrsv: !$
            \drawmatrix[width=.0]{b_i} \coloneqq
            \drawmatrix[lower]A^{-1}
            \drawmatrix[width=.0]{b_i}
        $!
!\drawskip!
    \end{lstlisting}
\end{experiment}

\begin{figure}[t]
    \includegraphics[width=\linewidth]{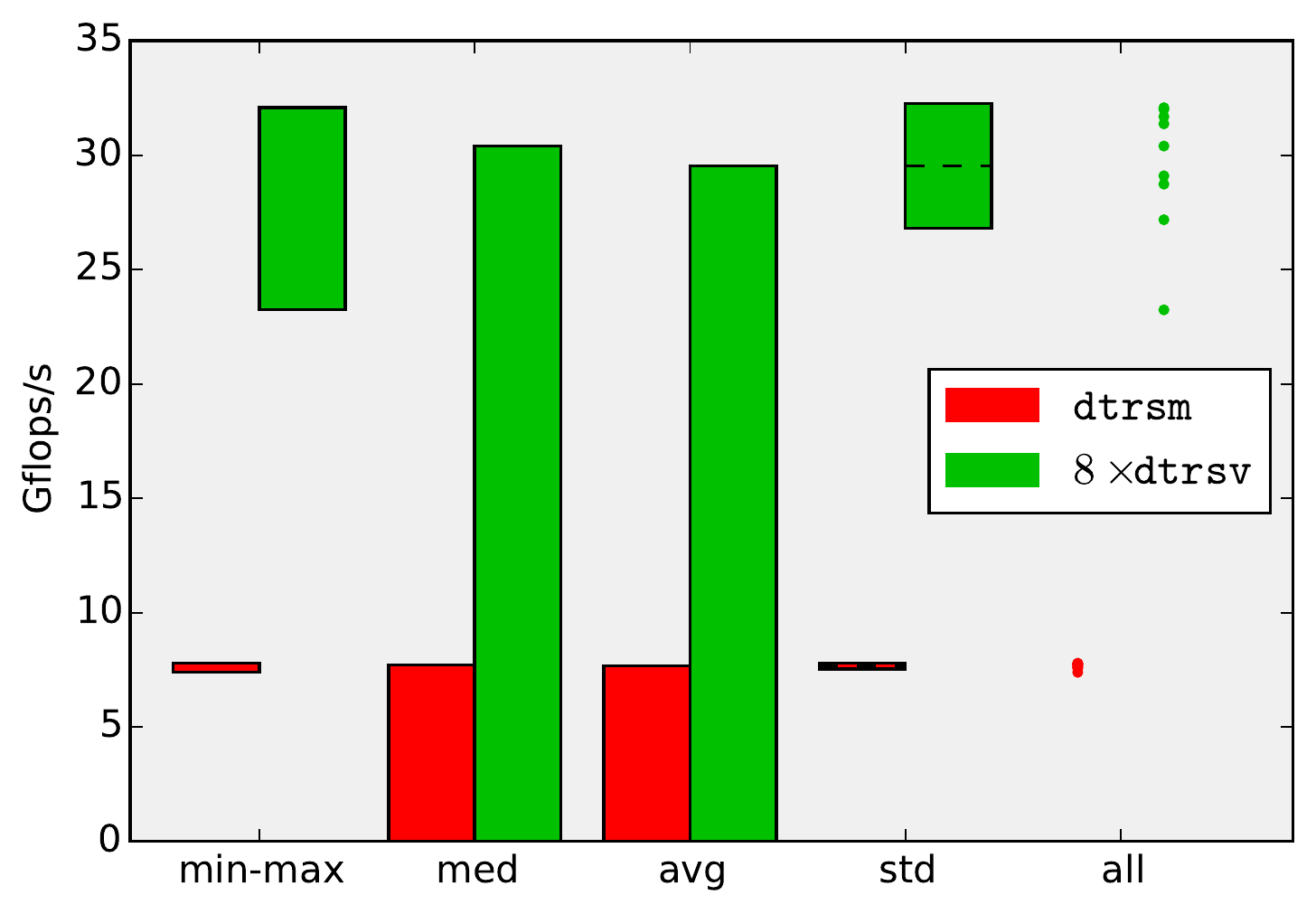}

    \caption{Performance of threaded {\tt dtrsm} vs. parallel {\tt dtrsv}'s}
    \label{fig:ex:parrange}
\end{figure}

The results in \autoref{fig:ex:parrange} suggest that the parallel {\tt
dtrsv}'s are considerably faster than the threaded {\tt dtrsm}'s, indicating
that  {\sc OpenBLAS} is not optimally parallelized for such extremely skewed
matrix sizes.


    \section{The {\sc\large ELAPS} Framework}
    \label{sec:elaps}
    \parsum{motivation and purpose}
The \elaps framework is built to support performance experiments combining the
features and scenarios described in \autoref{sec:experiment}.  In this section,
we present the structure of the framework, focusing on the aspects that make it
general and intuitive, yet powerful.

\begin{figure}
    \centering
    \begin{tikzpicture}[
        object/.style={
            draw,
            inner sep=5pt,
            fill=graybg
        },
        lib/.style={
            draw=gray, 
            fill opacity=.1,
            rounded corners
        },
        scale=1.36  
    ]
        \filldraw[lib, fill=plot3] (-1.1, 1.2) rectangle (5.1, -1.5);
        \node at (1, .8) {{\it Python}};
        \filldraw[lib, fill=plot2] (-1, .5) rectangle (5, -.5);
        \node at (2, 0) {\it PyQt4};
        \filldraw[lib, fill=plot1] (3.1, 1.1) rectangle (4.9, -1.4);
        \node at (4, .8) {{\it matplotlib}};
        \filldraw[lib, fill=plot4] (-1.1, .-1.6) rectangle (5.1, -2.4);
        \node at (3, -2) {{\it C/C++}};

        \node[object] (playmat) at (0, 0) {\playmat};
        \node[object] (viewer) at (4, 0) {\viewer};
        \node[object] (experiment) at (0, -1) {{\tt Experiment}};
        \node[object] (report) at (2, -1) {{\tt Report}};
        \node[object] (plot) at (4, -1) {{\tt plot}};
        \node[object] (sampler) at (1, -2) {\sampler};

        \draw[->, dashed] (playmat) -- (experiment);
        \draw[->] (experiment) to[out=270, in=180] (sampler);
        \draw[->] (sampler) to[out=0, in=270] (report);
        \draw[->] (report) -- (plot);
        \draw[->, dashed] (viewer) -- (report);
        \draw[->, dashed] (viewer) -- (plot);
    \end{tikzpicture}

    \caption{Structure of the \elaps framework}
    \label{fig:elaps}
\end{figure}
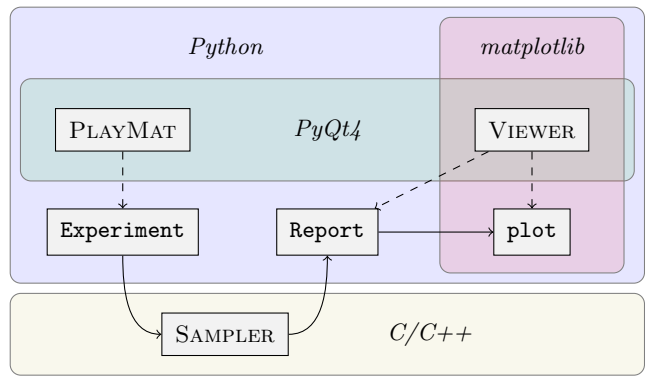

\parsum{structure}
As shown in \autoref{fig:elaps}, \elaps is structured in three layers.
\begin{itemize}
    \item The first, ``bottom'' layer (\autoref{sec:sampler}) is written in
        C/C++ and contains the \sampler, a low-level command line tool
        responsible for executing and timing individual kernels.  The \sampler
        has to be compiled for each specific combination of hardware and
        libraries (the only stage in which the user needs to configure the
        system); \elaps can interface with any number of \sampler{}s.
    \item The second, ``middle'' layer (\autoref{sec:elapslib}) is the Python
        library {\tt elaps}, which centers around the class {\tt Experiment}
        that implements the previously introduced experiments.  An {\tt
        Experiment} can be executed on different \sampler{}s, both locally or
        through job submission systems.  The outcome is a {\tt Report}, which
        provides not only structured access to the individual measurements, but
        also functionality to analyze different metrics and statistics.

        \parsum{plotting}
        This layer also includes the {\tt plot} module, which is based on the
        {\tt matplotlib} library, and is used to easily visualize {\tt Report}s
        in graphical form.\footnote{%
            All the plots in this paper were generated in this manner.
        }
    \item The third, ``top'' layer (\autoref{sec:gui}) adds a graphical user
        interface, written in {\sc PyQt4}, to both design {\tt Experiment}s in
        the \playmat and study {\tt Report}s and plots in the \viewer.
\end{itemize} 

The design of these three layers is discussed in the next subsections.


    \subsection{The {\sc\large Sampler}}
    \label{sec:sampler}
    \parsum{overview and workflow}
At the core of the \elaps framework is a low-level performance measurement tool
tailored to dense linear algebra operations: the \sampler.  This tool, earlier
versions of which were already utilized in \cite{modeling} and
\cite{tensorpred}, makes it possible to measure the performance of individual
kernel executions, implementing this work-flow:
\begin{enumerate}
    \item Read from {\tt stdin} a list of {\em calls}, i.e., kernel names with
        corresponding lists of arguments;
    \item execute the specified calls, thereby measuring their performance in
        terms of CPU cycles, and optionally through performance counters
        provided by  the {\sc Performance Application Programming Interface}
        (PAPI)~\cite{papi};
    \item print the measured performance numbers to the standard output.
\end{enumerate}

\parsum{"control flow" commands}
While reading the list of kernels from the standard input, the \sampler accepts
several special commands: {\tt go} executes, measures, and reports the results
of all the calls previously read; {\tt \{omp} and {\tt \}} respectively start
and end a list of calls to be executed as parallel {\sc OpenMP} tasks; {\tt
set\_counters} sets the PAPI counters for the next set of executions.

\parsum{Supported kernel format}
The \sampler accepts kernels and arguments in a format that agrees with the
conventions used by standard libraries such as BLAS and LAPACK: Each argument
is passed by reference, and is of type {\tt char~*}, {\tt int~*}, {\tt
float~*}, or {\tt double~*}.

In a dense linear algebra kernel, these arguments are of one of two types.
\begin{itemize}
    \item {\em Scalar arguments} point to scalar values, some of which may
        influence the kernel's behavior and control flow.  Examples include:
        flag arguments (e.g., {\tt side}, {\tt transA}), size arguments (e.g.,
        {\tt m}, {\tt n}), scalars (e.g., {\tt alpha}, {\tt beta}), and leading
        dimensions (e.g., {\tt ldA}, {\tt ldB}).

        Within the \sampler, scalar argument values are stored consecutively in
        an array.

    \item {\em Data arguments} point to memory regions that hold the
        mathematical objects (such as vectors or matrices) involved in the
        kernel call.  Generally\footnote{%
            Eigensolvers are a notable exception.
        }, the actual contents of these arguments do not affect the control
        flow; nonetheless, these arguments may still have a significant impact
        on performance, depending on their location in the memory hierarchy.

        The \sampler has two mechanisms to treat data arguments:
        \begin{itemize}
            \item {\em Named variables} are designated memory regions
                referenced by a variable names.  A set of features to allocate
                ({\tt{\it x}malloc}\footnote{%
                    $\mathtt x \in \{\mathtt i, \mathtt s, \mathtt d, \mathtt
                    c, \mathtt z\}$ identifies the data-type.
                }), compute offsets ({\tt{\it x}offset}) and free ({\tt free})
                such variables give users full control over where operands are
                stored in relation to each other.
            \item {\em Dynamic memory} offers a fast way to pass ``unnamed''
                memory regions as data arguments.  Within one call all such
                regions are guaranteed to be disjoint; across calls, however,
                the same memory regions may be reused arbitrarily.
        \end{itemize}

        To set up the contents of data arguments, the \sampler provides a set
        of simple utility-type kernels: {\tt{\it x}memset} fills every entry in
        a buffer with a single value, {\tt{\it x}gerand} fills it with random
        values (uniform in $]0, 1[$), and {\tt{\it x}porand} generates a random
        symmetric (or Hermitian) positive definite matrix.  Furthermore,
        {\tt{\it x}readfile} and {\tt{\it x}writefile}, read matrices from and
        write them to binary files, respectively.
\end{itemize}


    \subsection{The {\tt\large elaps} Package}
    \label{sec:elapslib}
    \parsum{repeat: {\tt Experiment} $\rightarrow$ {\tt Report}}
The middle layer of the \elaps framework centers around the experimental
features introduced in \autoref{sec:experiment}, encoded in the Python class
{\tt Experiment}.  Instances of this class form the starting point for
performance experiments; executing them using \sampler{}s ultimately leads to
{\tt Report}s, which can be analyzed with respect to a variety of metrics and
statistics. 

\subsubsection{{\tt Experiment}s}
\parsum{static portability and functionality}
{\tt Experiment} instances are both a static description of experiments, which
are easily stored to and loaded from strings and files for portability, but
also feature functionality to support their design and handling.

\parsum{signatures}
The kernel configurations at the center of each {\tt Experiment} are its
connection to libraries such as BLAS or LAPACK.  While the interfaces of such
libraries aim at being general by accommodating multiple functionalities,
precisely because of their generality they are often unintuitive and hard to
memorize.  To counter this problem, {\tt elaps} uses optional ``{\tt
Signature}s'' to annotate kernels, thereby providing possible value ranges and
semantic connections between arguments.  In the end, these {\tt Signature}s
allow {\tt Experiment}s to expose feasible values for arguments (such as {\tt
trans} or {\tt uplo}) and automatically derive connected arguments such as
operand sizes and leading dimensions, both within a single kernel and across
multiple kernels.

\parsum{submitting}
The execution of an {\tt Experiment} is initiated by the {\tt submit} method.
This method first generates the sequence of kernel calls for the \sampler and a
shell script for its execution; it then either executes this script is locally
or submits it to a batch job system.

\subsubsection{Execution on \sampler{}s}
In this section, we describe how the {\tt Experiment} features are translated
into commands for the \sampler.

\parsum{ranges}
As input, the \sampler expects a raw list of kernel invocations.  To produce
this list, all ranges and repetitions in an {\tt Experiment} are completely
unrolled, thereby evaluating any symbolic (range-dependent) variable.  The {\sc
OpenMP}-range is translated directly to the \sampler's {\tt\{omp} and {\tt\}}
commands.  Using the parameter-range to vary the number of library threads
requires to interface with said library;  to avoid library-dependent kernels
and \sampler features, we do so through environment variables (e.g., {\tt
OPENBLAS\_NUM\_THREADS}) and by starting the sampler separately for each thread
count.

\parsum{data and varying}
Data arguments in kernels are allocated as named variables in the \sampler at
the beginning of the input.  Arguments that vary with repetitions or the
sum/{\sc OpenMP}-range (i.e., they point to different locations) are first
allocated as a single large block and then subdivided by calculating
appropriate offsets, resulting in individual variables for each repetition and
range iteration.

\parsum{PAPI}
Finally, PAPI counters are also set at the beginning through the {\tt
set\_counters} command.

\subsubsection{Reports}
\parsum{structured data}
Each {\tt Experiment} execution results in a report file that, when read into
{\tt elaps}, turns into a {\tt Report} instance.  This object serves as a
structured representation of the obtained measurements with respect to the
underlying {\tt Experiement}:  Raw measurements are accessed through the
hierarchy ``parameter-range value $\rightarrow$ repetition $\rightarrow$
sum/{\sc OpenMP}-range value $\rightarrow$ kernel'' and yield the cycle count
or PAPI counter measurements.  Separately, a ``reduced'' view on the results
accumulates the sum/{\sc OpenMP}-range and the kernels according to the
experiment semantic.

\parsum{metrics}
To turn these structured yet raw measurement results into more meaningful
quantities, {\em metrics} combine them with the kernels' flop counts and
information on the their execution environment.  The easily extensible set of
metrics ranges from ``execution time in seconds'' to ``\si{Gflops/s}'' and
``efficiency''.

\parsum{statistics}
While a metric converts measurements values one-by-one, results from multiple
repetitions are combined by statistics, such as ``minimum'', ``maximum'',
``median'' or ``standard deviation''.  As motivated in
\autoref{sec:experiment}, the results from first repetitions are optionally
discarded to hide overhead effects and make statistics more representative of
in-application invocations.

\subsubsection{Plotting}
\elaps's {\tt plot} module generates {\tt matplotlib} figures from the
structured data in {\tt Report}s under consideration of both metrics and
statistics.  Depending on the type of experiment, it automatically generates
appropriate bar- or line-plots that are easily exported to various file
formats.


    \subsection{The Graphical User Interface}
    \label{sec:gui}
    \parsum{motivation for GUI}
\elaps's {\tt Experiment} and {\tt Report} Python classes establish a flexible
and powerful foundation for performance evaluations at a scripting level.  In
order to enable performance experimentation in an explorative fashion,  and to
facility more intuitive interaction, \elaps features a graphical user
interface.  As shown in \autoref{fig:elaps}, this interface consists of
components: While the \playmat serves as a ``playing mat'' to develop {\tt
Experiment}s, the \viewer interactively visualizes {\tt Report}s.

\begin{figure}[t]
    \includegraphics[width=\linewidth]{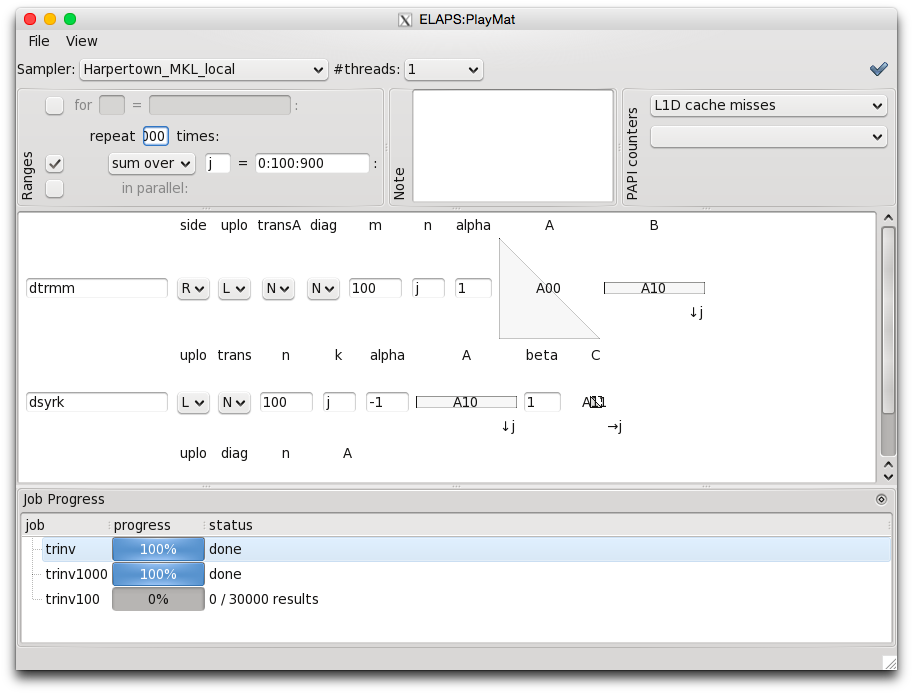}

    \caption{The \playmat through {\sc X11}.}
    \label{fig:playmat}
\end{figure}

\parsum{\playmat}
The \playmat, shown in \autoref{fig:playmat}, allows interactive access to the
full functionality of an {\tt Experiment}.  To further guide the user, among
other things, it visualizes the {\tt Experiment}'s kernels based on their {\tt
Signature}s similar to how they are presented in this paper and automatically
calculates matrix sizes and (if desired) deducible arguments, such as leading
dimensions.  It furthermore provides progress tracking of executing {\tt
Experiment}s and can load completed {\tt Report}s directly in the \viewer.

\begin{figure}[t]
    \includegraphics[width=\linewidth]{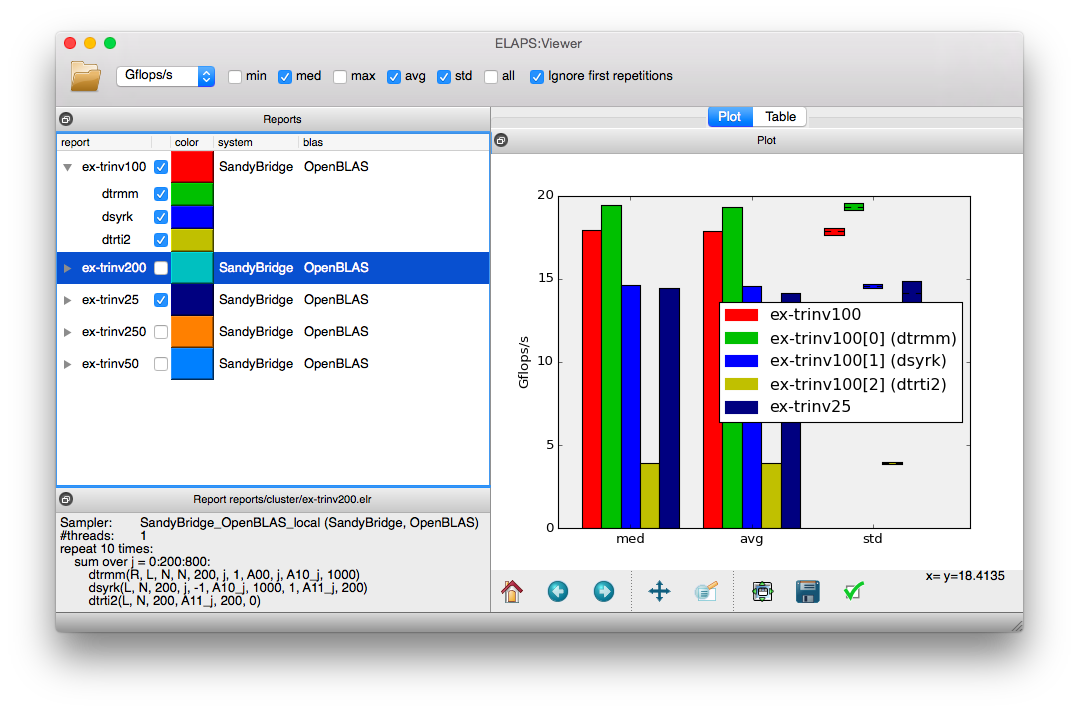}

    \caption{The \viewer on {\sc Mac~OS~X}}
    \label{fig:viewer}
\end{figure}

\parsum{\viewer}
The \viewer, shown in \autoref{fig:viewer}, is an interactive means to analyze
and compare {\tt Report}s. It provides all metrics applicable to the selected
{\tt Report}s and its statistical plots are easily manipulated and exported.


    \section{Application Examples}
    \label{sec:examples}
    \parsum{Application examples}
In this section, we demonstrate the use of the \elaps framework in several
application examples.  For this purpose, we deliberately chose a wide range of
hardware systems and kernel libraries.


    \subsection{Algorithm Selection: Tensor Contractions}
    \label{sec:tensors}
    \newcommand\drawtensor[1]{

        \draw[densely dotted, gray]
            (0, 0, 1) -- (1, 0, 1)  
            (0, 0, 1) -- (0, 1, 1)  
            (0, 0, 0) -- (0, 0, 1); 
        \node (X) at (.5, .5, .5) {#1};
        \draw[gray]
            (0, 0, 0) -- (1, 0, 0)  
            (0, 1, 0) -- (1, 1, 0)
            (0, 1, 1) -- (1, 1, 1)
            (0, 0, 0) -- (0, 1, 0)  
            (1, 0, 0) -- (1, 1, 0)
            (1, 0, 1) -- (1, 1, 1)
            (1, 0, 0) -- (1, 0, 1)  
            (0, 1, 0) -- (0, 1, 1)
            (1, 1, 0) -- (1, 1, 1);
}

\parsum{problem description and algorithms}
Let us consider the tensor contraction (in Einstein notation)
\begin{equation}
    C_{abc} \coloneqq A_{ak} B_{kcb}
    \label{eqn:tensor}
\end{equation}
with 
$A \in \mathbb R^{\num{1250} \times \num{750}}$, 
$B \in \mathbb R^{\num{750} \times \num{500} \times n}$, and 
$C \in \mathbb R^{\num{1250} \times n \times \num{500}}$, where $n$ is between
\num{100} and \num{10000}.  Using an explicit index notation, $C$ is computed as
\begin{equation}
    \forall a, \forall b, \forall c .\  C[a, b, c]
    \coloneqq \sum\limits_{k} A[a, k] \cdot B[k, c, b]
\end{equation}
and can be visualized as follows:
\begin{center}
    \begin{tabular}{cc}
        $
            \begin{tikzpicture}[y=1.25cm, x=.1cm, z={(45:.354cm)}, baseline=(X.base), scale=.8]
                \drawtensor{$C$}
                \node[gray, anchor=east] at (0, .5, 0) {$_a$};
                \node[gray, anchor=north] at (.5, 0, 0) {$_b$};
                \node[gray, anchor=north west, inner sep=0] at (1, 0, .5) {$_c$};
            \end{tikzpicture}
            \coloneqq
            \begin{tikzpicture}[y=1.25cm, x=.75cm, z=0cm, baseline=(X.base), scale=.8]
                \drawtensor{$A$}
                \node[gray, anchor=east] at (0, .5) {$_a$};
                \node[gray, anchor=north] at (.5, 0) {$_k$};
            \end{tikzpicture}
            \;
            \begin{tikzpicture}[y=.75cm, x=.5cm, z={(45:.071cm)}, baseline=(X.base), scale=.8]
                \drawtensor{$B$}
                \node[gray, anchor=east] at (0, .5, 0) {$_k$};
                \node[gray, anchor=north] at (.5, 0, 0) {$_c$};
                \node[gray, anchor=north west, inner sep=0] at (1, 0, .5) {$_b$};
            \end{tikzpicture}
        $ &$
            \begin{tikzpicture}[y=1.25cm, x=1cm, z={(45:.354cm)}, baseline=(X.base), scale=.8]
                \drawtensor{$C$}
                \node[gray, anchor=east] at (0, .5, 0) {$_a$};
                \node[gray, anchor=north] at (.5, 0, 0) {$_b$};
                \node[gray, anchor=north west, inner sep=0] at (1, 0, .5) {$_c$};
            \end{tikzpicture}
            \coloneqq
            \begin{tikzpicture}[y=1.25cm, x=.75cm, z=0cm, baseline=(X.base), scale=.8]
                \drawtensor{$A$}
                \node[gray, anchor=east] at (0, .5) {$_a$};
                \node[gray, anchor=north] at (.5, 0) {$_k$};
            \end{tikzpicture}
            \;
            \begin{tikzpicture}[y=.75cm, x=.5cm, z={(45:.707cm)}, baseline=(X.base), scale=.8]
                \drawtensor{$B$}
                \node[gray, anchor=east] at (0, .5, 0) {$_k$};
                \node[gray, anchor=north] at (.5, 0, 0) {$_c$};
                \node[gray, anchor=north west, inner sep=0] at (1, 0, .5) {$_b$};
            \end{tikzpicture}
        $ \\
        ($n = \num{100}$) &($n = \num{1000}$)
    \end{tabular}
\end{center}

\parsum{implementations and experiments}
A natural approach to efficiently compute such tensor contractions is to
utilize the highly optimized {\tt dgemm} kernel.  For Contraction
(\ref{eqn:tensor}), there are two ways of casting the computation as a series
of {\tt dgemm}'s:

\begin{equation}
    \forall b \in \{1, \ldots, n\} : \quad 
    \begin{tikzpicture}[y=1.25cm, x=1cm, z={(45:.354cm)}, baseline=(X.base)]
        \filldraw[plot3, fill opacity=.1] (.5, 0, 0) -- (.5, 1, 0) -- (.5, 1, 1) -- (.5, 0, 1) -- (.5, 0, 0);
        \drawtensor{$C$}
        \draw[plot3, ->] (0, 0, 0) -- (1, 0, 0);
        \node[anchor=north] at (.5, 0, 0) {$b$};
    \end{tikzpicture}
    \coloneqq
    \begin{tikzpicture}[y=1.25cm, x=.75cm, z=0cm, baseline=(X.base)]
        \fill[plot3, fill opacity=.1] (0, 0) rectangle (1, 1);
        \drawtensor{$A$}
        \draw[plot3, fill opacity=.1] (0, 0) rectangle (1, 1);
    \end{tikzpicture}
    \;
    \begin{tikzpicture}[y=.75cm, x=.5cm, z={(45:.707cm)}, baseline=(X.base)]
        \filldraw[plot3, fill opacity=.1] (0, 0, .5) rectangle (1, 1, .5);
        \drawtensor{$B$}
        \draw[plot3, ->] (1, 0, 0) -- (1, 0, 1);
        \node[anchor=north west, inner sep=0] at (1, 0, .5) {$b$};
    \end{tikzpicture}
    \label{eqn:tensor:forb}
\end{equation}

\begin{equation}
    \forall c \in \{1, \ldots, \num{500}\} : \quad 
    \begin{tikzpicture}[y=1.25cm, x=1cm, z={(45:.354cm)}, baseline=(X.base)]
        \filldraw[plot3, fill opacity=.1] (0, 0, .5) rectangle (1, 1, .5);
        \drawtensor{$C$}
        \draw[plot3, ->] (1, 0, 0) -- (1, 0, 1);
        \node[anchor=north west, inner sep=0] at (1, 0, .5) {$c$};
    \end{tikzpicture}
    \coloneqq
    \begin{tikzpicture}[y=1.25cm, x=.75cm, z=0cm, baseline=(X.base)]
        \fill[plot3, fill opacity=.1] (0, 0) rectangle (1, 1);
        \drawtensor{$A$}
        \draw[plot3, fill opacity=.1] (0, 0) rectangle (1, 1);
    \end{tikzpicture}
    \;
    \begin{tikzpicture}[y=.75cm, x=.5cm, z={(45:.707cm)}, baseline=(X.base)]
        \filldraw[plot3, fill opacity=.1] (.5, 0, 0) -- (.5, 1, 0) -- (.5, 1, 1) -- (.5, 0, 1) -- (.5, 0, 0);
        \drawtensor{$B$}
        \draw[plot3, ->] (0, 0, 0) -- (1, 0, 0);
        \node[anchor=north] at (.5, 0, 0) {$c$};
    \end{tikzpicture}
    \label{eqn:tensor:forb}
\end{equation}

\parsum{experiment design}
An inspection of these algorithms reveals that they both execute a {\tt dgemm}
of fixed size on varying data; in particular, the number of invocations in
Algs.~$\forall b$ and $\forall c$ are, respectively, $n$ and \num{500}.  By
virtue of this observation, Experiments~10 and 11 only perform \num{10}
repetitions, thus reducing the experimentation time; while the results will not
be meaningful estimates for the total execution time, they will expose the same
computational efficiency (expressed in \si{Gflops/s}) as the full algorithms.
Furthermore, since Alg.~$\forall b$ operates on matrices of fixed size and
independent of $n$, in Experiment~10 we also  avoid the use of the parameter
range.

\begin{experiment}{Tensor algorithm $\forall b$}{PowerPC A2}{ESSL}{$
    A \in \mathbb R^{\num{1250} \times \num{750}},
    B_{\rm \downarrow rep} \in \mathbb R^{\num{750} \times \num{500}},
    C_{\rm rep} \in \mathbb R^{\num{1250} \times \num{500}}
$}
    \drawmatrixset{bbox style={fill=white, draw=gray, opacity=.5}}%
    \begin{lstlisting}
#threads = 64
repeat 10 times:
    dgemm: !$
        \drawmatrix[width=.4]{C_{\rm rep}} \coloneqq 
        \drawmatrix[width=.6] A \ 
        \drawmatrix[height=.6, width=.4]{B_{\rm \downarrow rep}}
    $!
!\drawskip!
    \end{lstlisting}
\end{experiment}

\begin{experiment}{Tensor algorithm $\forall c$}{PowerPC A2}{ESSL}{$
    A \in \mathbb R^{\num{1250} \times \num{750}},
    B_{\rm rep} \in \mathbb R^{\num{750} \times n},
    C_{\rm \downarrow rep} \in \mathbb R^{\num{1250} \times n}
$}
    \drawmatrixset{bbox style={fill=white, draw=gray, opacity=.5}}%
    \begin{lstlisting}
#threads = 64
for $n$ = 100:10:1000:
    repeat 10 times:
        dgemm: !$
            \drawmatrix[bbox width=.8, width=.08]{C_{\rm \downarrow rep}} \coloneqq 
            \drawmatrix[width=.6] A \ 
            \drawmatrix[height=.6, bbox width=.8, width=.08]{B_{\rm rep}}
        $!
!\drawskip!
    \end{lstlisting}
\end{experiment}

\parsum{system setup}
We perform Experiments~10 and 11 on a 16-core {\sc IBM PowerPC~A2} node of the
{\sc IBM BlueGene} installation {\sc JUQUEEN} at the {\sc Jülich Supercomputing
Center}, linked to IBM's optimized ESSL library.  Only the \sampler is executed
on this compute node, running the lightweight CNK operating system; it is
accessed by {\tt elaps} from a {\sc Red Hat Enterprise Linux}~6.6 front-end
node through the {\sc LoadLeveler} batch job system.

\begin{figure}[t]
    \includegraphics[width=\linewidth]{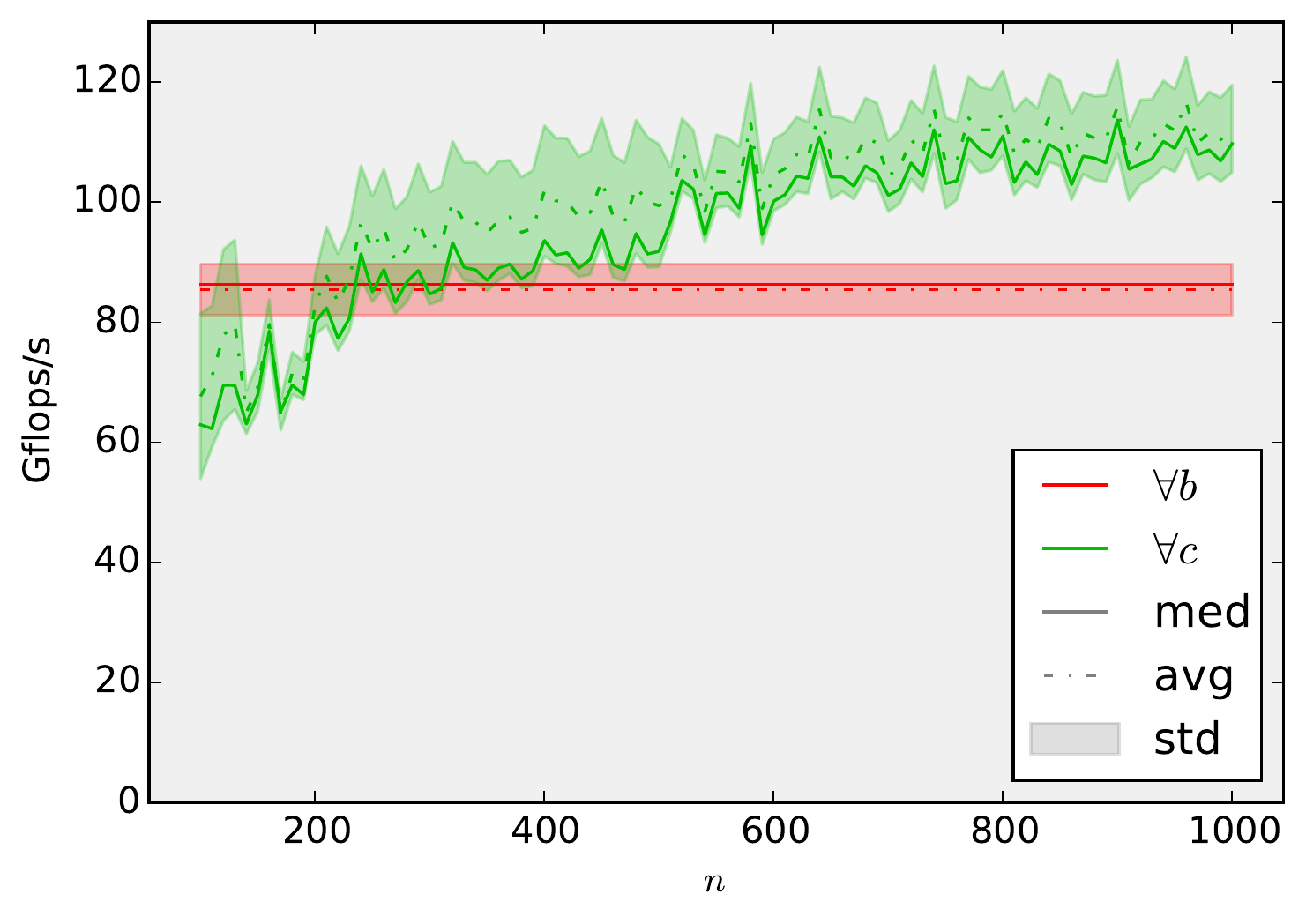}

    \caption{Comparison of {\tt dgemm}-based algorithms for tensor contraction
    \ref{eqn:tensor}}
    \label{fig:tensors}
\end{figure}


\parsum{results}
\autoref{fig:tensors} suggests that neither of the two algorithms is optimal
for all cases:  While for a small dimension $n$, algorithm $\forall b$ is
better, algorithm $\forall c$ dominates for large $n$.  Interestingly, the
crossover point is not at $n = \num{500}$, where both algorithms work with
matrices of equal size, but already around $n = \num{300}$. 


    \subsection{Library Selection: Sylvester Equation}
    \label{sec:sylv}
    \parsum{motivation, problem and setup}
Choosing the right library can be a crucial step in attaining high performance.
In this section, we demonstrate its importance by considering the triangular
Sylvester equation
\begin{equation}
    \drawmatrix[upper]A \ \drawmatrix[width=.666]X + 
    \drawmatrix[width=.666]X \ \drawmatrix[upper, size=.666]B =
    \drawmatrix[width=.666]C,
    \label{eqn:sylv}
\end{equation}
to be solved for $X$, which is central to problems in control theory.  In
addition to LAPACK's {\tt dtrsyl}, several other libraries offer this kernel
with the same interface.  In our tests, we consider
\begin{itemize}
    \item LAPACK~3.5~\cite{lapack} linked to {\sc
        OpenBLAS}~0.2.14~\cite{openblas},
    \item RECSY~0.01~\cite{recsy} linked to {\sc OpenBLAS},
    \item {\sc libFLAME}\footnote{%
            While {\sc libFLAME}'s LAPACK interface does by default not call
            its optimized Sylvester solver, it is easily exposed.
        }~5.1.0-18~\cite{flame} linked to {\sc OpenBLAS}, and
    \item {\sc Intel}'s commercial MKL~11.0~\cite{mkl}.
\end{itemize}

\parsum{system setup}
To compare these libraries, we launch Experiment 12 on a 10-core {\sc Intel
IvyBridge E5-2680~v2} processor with \sampler{}s linked to the above libraries.
This machine, which is part of a compute cluster is accessed through the {\sc
Platform LSF}~9.1.2 batch job system with both the front-end ({\tt elaps}) and
back-end (the \sampler{}s) running {\sc Scientific Linux}~6.6.

\begin{experiment}{Sylvester solver}{E5-2680 v2}{\normalfont\it see above}{$
    A \in \mathbb R^{i \times i},
    B \in \mathbb R^{i / 2 \times i / 2},
    C \in \mathbb R^{i \times i / 2}
$}
    \begin{lstlisting}
#threads = 10
for $i$ = 100:100:6000:
    repeat 10 times:
        dtrsyl: !$
            {
                \drawmatrixset{bbox style={fill=white, draw=gray, opacity=.5}}
                \drawmatrix[bbox size=1, bbox width=.5, size=.01, width=.005]C 
            }
            \coloneqq {\rm Sylv}\left( 
            \drawmatrix[upper, size=1, opacity=.5]A,
            \drawmatrix[upper, size=.5, opacity=.5]B,
            {
                \drawmatrixset{bbox style={fill=white, draw=gray, opacity=.5}}
                \drawmatrix[bbox size=1, bbox width=.5, size=.01, width=.005]C 
            }
            \right)
        $!
!\drawskip!
    \end{lstlisting}
\end{experiment}

\begin{figure}[t]
    \includegraphics[width=\linewidth]{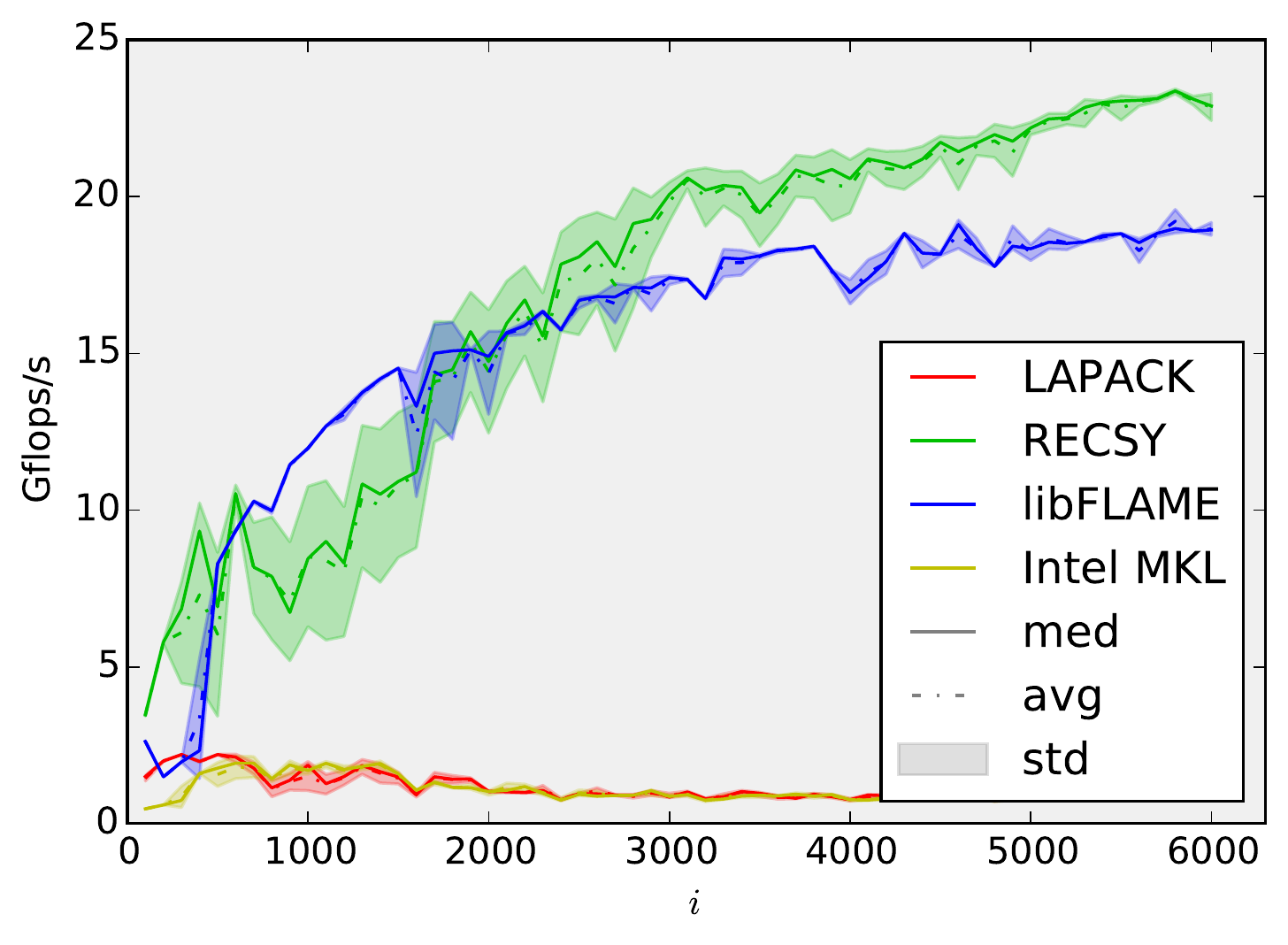}

    \caption{Comparison of libraries for the triangular Sylvester equation}
    \label{fig:sylv}
\end{figure}

\parsum{results}
\autoref{fig:sylv} shows the performance attained by the different libraries.
LAPACK, which only provides an unblocked algorithm for {\tt dtrsyl}, reaches
\SI{2}{Gflops/s} for small problems but eventually falls below
\SI{1}{Gflops/s}. The specialized RECSY library on the other hand attains the
best performance of up to \SI{24.5}{Gflops/s}.  {\sc libFLAME} is initially
competitive with RECSY but eventually tops at \SI{20.5}{Gflops/s}.
Surprisingly, the otherwise very efficient MKL seems poorly optimized for this
problem and is as fast as LAPACK. 


    \subsection{Multithreading: Sequence of LUs}
    \label{sec:lus}
    \parsum{origin and alternatives}
In a certain type of electronic structure calculations~\cite{lus}, one has to
solve a series of fairly small linear systems. As already mentioned, a possible
approach involves the LU factorization ({\tt dgetrf}) and two triangular linear
systems ({\tt dtrsm}) for each matrix; since each system only involves two
right-hand sides, the cost for the {\tt dtrsm}'s is entirely negligible.

On multi-core machines, the sequence of LUs can be parallelized at the
granularity of one matrix, via a multi-threaded {\tt dgetrf} kernel, or by
assigning different matrices to different threads, via {\sc OpenMP} (hybrid
solutions are also possible).  Performance-wise, there are arguments both in
favor and against each of these two alternatives:  Using BLAS's internal
parallelism ensures that only one kernel uses the CPU's caches at a time; on
the other hand, {\sc OpenMP}'s parallelism increases the amount of work that
the CPU's cores can perform simultaneously. 

\parsum{Experiment setup}
The next three experiments use \elaps's sum-range and {\sc OpenMP}-range
constructs to model the scenarios in which 1) a multi-threaded kernel is used,
2) {\sc OpenMP} runs sequential kernels in parallel, and 3) {\sc OpenMP} runs
multi-threaded kernels in parallel.\footnote{This third experiment is not
displayed: It is obtained from Experiment 14, changing {\tt \#threads} to 8.}
Each of them measures the time it takes to LU decompose an increasing number of
square matrices of size \num{800}.

\parsum{system setup}
These three experiments are executed on a {\sc MacBook Pro} running OS~X~10.9.4
with a quad-core {\sc Intel Haswell i7-4850HQ} CPU (Turbo Boost disabled) using
{\sc Apple}'s {\sc Accelerate} framework; both the \sampler and {\tt elaps} run
on the same platform. 

\begin{experiment}{Threaded LUs}{i7-4850HQ}{Accelerate}{$
    A_i \in \mathbb R^{\num{800} \times \num{800}}
$}
    \drawmatrixset{bbox style={fill=white, draw=gray, opacity=.5}}%
    \begin{lstlisting}
#threads = 8
for $n$ = 1:50:
    repeat 10 times:
        sum over $i$ = 1:$n$:
            dgetrf: !$
                \drawmatrix{A_i} = {\rm LU}\left(\drawmatrix{A_i}\right) 
            $!
!\drawskip!
    \end{lstlisting}
\end{experiment}

\begin{experiment}{Parallel LUs}{i7-4850HQ}{Accelerate}{$
    A_i \in \mathbb R^{\num{800} \times \num{800}}
$}
    \drawmatrixset{bbox style={fill=white, draw=gray, opacity=.5}}%
    \begin{lstlisting}
#threads = 1
for $n$ = 1:50:
    repeat 10 times:
        in parallel $i$ = 1:$n$:
            dgetrf: !$
                \drawmatrix{A_i} = {\rm LU}\left(\drawmatrix{A_i}\right) 
            $!
!\drawskip!
    \end{lstlisting}
\end{experiment}

\begin{figure}[t]
    \includegraphics[width=\linewidth]{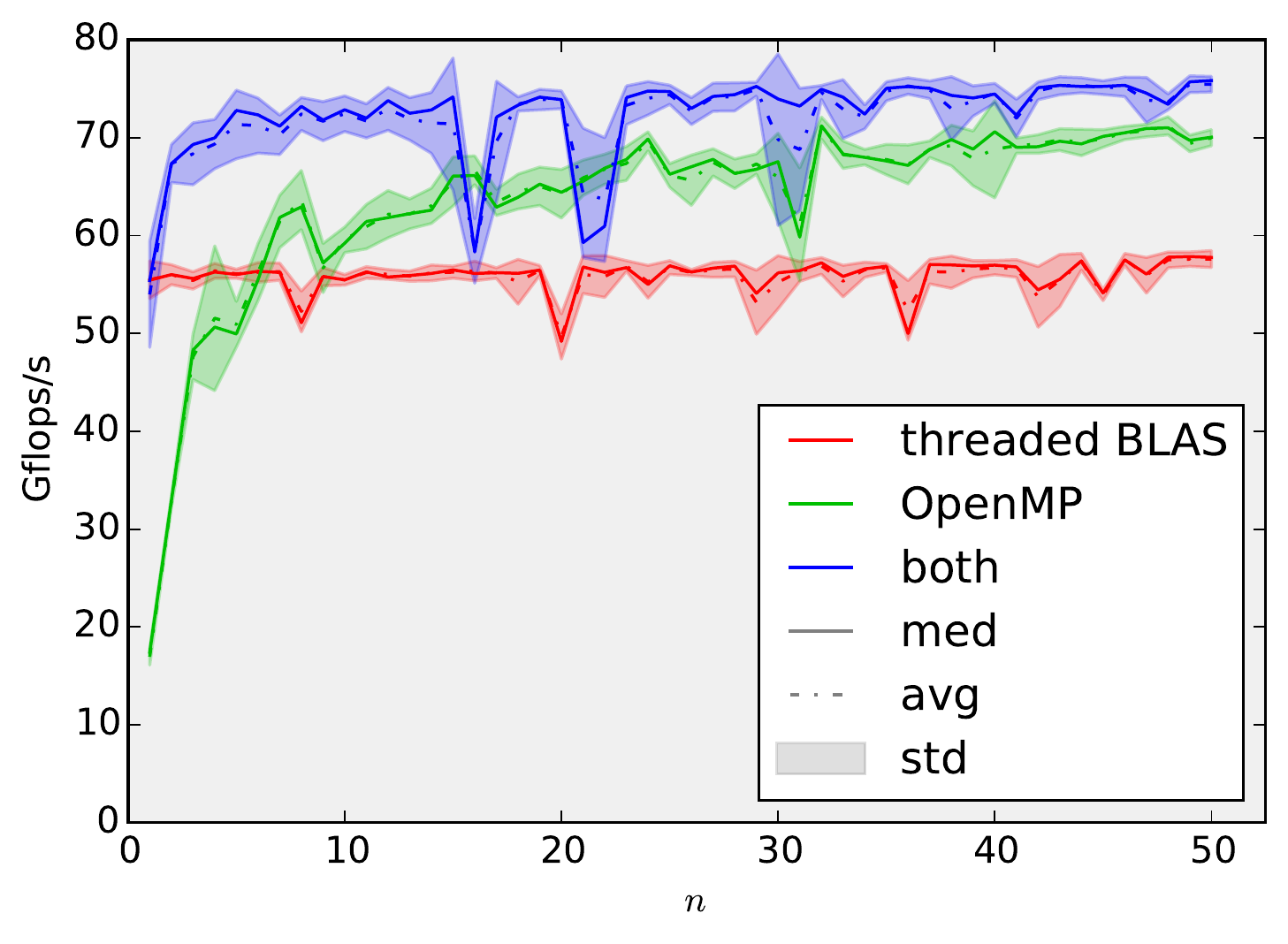}

    \caption{Multi-threading paradigms for a sequence of
    LUs\protect\footnotemark}
    \label{fig:lus}
\end{figure}

\footnotetext{
  Even though the Turbo Boost was disabled and no other
  application was running, 
  noticeable performance fluctuations were observed. 
  This is to be expected on laptop-systems.  
}


\parsum{results}
\autoref{fig:lus} indicates that of the two ``pure'' approaches, if more than
\num{8} LU's are performed (i.e., more than the CPU has hardware threads), {\sc
OpenMP} with single-threaded kernels outperforms {\sc Accelerate}'s parallel
kernel.  However, the mixed approach, in which {\sc Accelerate} uses up to
\num{8} threads, while {\sc OpenMP} is allowed to schedule the LU decomposition
tasks, is even more efficient, reaching up to \SI{75}{Gflops/s}.


    \subsection{Algorithmic Optimization: GWAS}
    \label{sec:gwas}
    
\parsum{problem description}
Genome Wide Association Studies (GWAS) investigate how human traits (e.g. eye
color or genetic deceases) are related to certain locations in the human
genome~\cite{gwas1, gwas2}.  Computationally, GWAS can be cast as a sequence of
Generalized Least Squares (GLS) problems
\begin{equation}
    \drawmatrix[height=.1, width=0]{b_i} \coloneqq
    \left(\drawmatrix[height=.1]{X_i^T} \ \drawmatrix M^{-1}
    \drawmatrix[width=.1]{X_i}\right)^{-1} \drawmatrix[height=.1]{X_i^T} \
    \drawmatrix M^{-1} \drawmatrix[width=0]y,
    \label{eqn:gwas}
\end{equation}
where $M \in \mathbb R^{n \times n}$ is symmetric positive definite, $X_i \in
\mathbb R^{n \times p}$, $y \in \mathbb R^n$, and $b_i \in \mathbb R^{p}$ with
$\num{1000} \leq n \leq \num{5000}$, $p \leq 20$ and $i \in \{1, \ldots m\}$,
where $m$ can be in the millions.

\parsum{"stupid" algorithm}
A straightforward implementation of this equation (e.g., using R or Matlab)
might compute each $b_i$ individually by solving \autoref{eqn:gwas} from right
to left, as modeled in this next experiment (with $n = \num{1000}$ and $p =
\num{4}$):

\begin{experiment}{Multiple GLS}{Xeon Phi}{MKL}{$
        M \in \mathbb R^{\num{1000} \times \num{1000}},
        y \in \mathbb R^{\num{1000}},
        b \in \mathbb R^4,
        X_i \in \mathbb R^{\num{1000} \times \num{4}},
        S \in \mathbb R^{\num{4} \times \num{4}}
$}
    \drawmatrixset{bbox style={fill=white, draw=gray, opacity=.5}}%
    \begin{lstlisting}
#threads = 240
for $m$ = 100:100:1000:
    sum over $i$ = 1:$m$:
        dposv!\footnotemark!: !$
            \drawmatrix[width=0]y \coloneqq 
            \drawmatrix M^{-1} \drawmatrix[width=0]y
        $!
        dgemv!\footnotemark!: !$
            \drawmatrix[height=.1, width=0]b \coloneqq
            \drawmatrix[height=.1]{X_i^T}^{-1} \drawmatrix[width=0]y
        $!
        dpotrs!\footnotemark!: !$
            \drawmatrix[width=.1]{X_i} \coloneqq 
            \drawmatrix{M}^{-1} \drawmatrix[width=.1]{X_i}
        $!
        dsyrk: !$
            \drawmatrix[size=.1] S \coloneqq
            \drawmatrix[height=.1]{X_i^T} \drawmatrix[width=.1]{X_i}
        $!
        dposv: !$
            \drawmatrix[height=.1, width=0] b \coloneqq
            \drawmatrix[size=.1] S^{-1} \drawmatrix[height=.1, width=0]b
        $!
!\drawskip!
    \end{lstlisting}
\end{experiment}
\addtocounter{footnote}{-3}
\stepcounter{footnote}\footnotetext{%
    {\tt dposv}: Cholesky decomposition + linear system solve.
}
\stepcounter{footnote}\footnotetext{%
    {\tt dgemv}: General matrix vector product.
}
\stepcounter{footnote}\footnotetext{%
    {\tt dpotrs}: Linear system solve following a Cholesky decomposition.
}

\parsum{system configuration}
For Experiments 15 and 16, we choose a 60-core {\sc Intel Xeon Phi}
co-processor using {\sc Intel} MKL.  \elaps's python library and the \playmat
are  run on this system's Host processor ({\sc Scientific Linux}~6.6); only the
\sampler is executed natively on the co-processor.

\begin{figure}[t]
    \includegraphics[width=\linewidth]{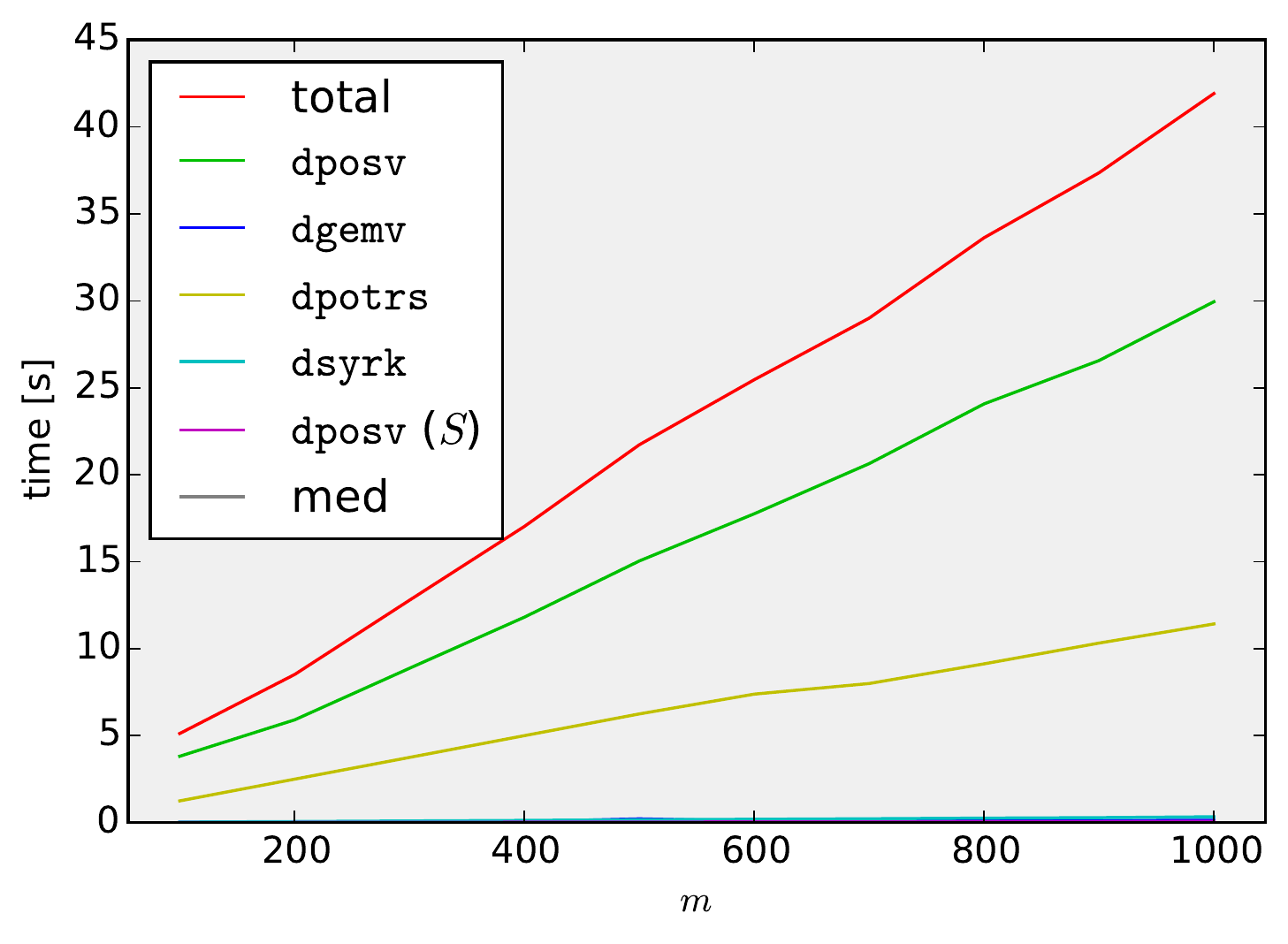}

    \caption{Timing breakdown for a sequence of GLS solves}
    \label{fig:gwas}
\end{figure}

\parsum{results}
\autoref{fig:gwas} shows both the execution time of Experiment 14, as well as a
breakdown thereof. The runtime is clearly dominated by the {\tt dposv} and {\tt
dpotrs} kernels involving $M$.

\parsum{Move loop-independent {\tt dposv} out}
From a first analysis of these two kernels, one realizes that the {\tt dposv}
($y \coloneqq M^{-1} y$) is independent of $i$, and can thus be taken out of
the loop and computed just once. This modification reduces {\tt dposv}'s
contribution to the total runtime by a factor of $m$, effectively shifting the
bottleneck onto {\tt dpotrs}.

\parsum{Blocking}
A further analysis reveals that all the {\tt dpotrs} linear systems involve the
same matrix $M$; this observation suggests to combine the right-hand sides $X_i
\in \mathbb R^{m \times \num{4}}$ for all $m$ iterations into a single large
matrix $X \in \mathbb R^{\num{1000} \times \num{4} m}$.  The following
experiment solves a linear system with this matrix as the right-hand side with
a single invocation of {\tt dpotrs}:

\begin{experiment}{Combining $X_i$}{Xeon Phi}{MKL}{$
        M \in \mathbb R^{\num{1000} \times \num{1000}},
        X \in \mathbb R^{\num{1000} \times \num{4} m}
$}
    \drawmatrixset{bbox style={fill=white, draw=gray, opacity=.5}}%
    \begin{lstlisting}
#threads = 240
for $m$ = 100:100:1000:
    dpotrs: !$
        \drawmatrix[height=.25, bbox width=1, width=.25]X
        \coloneqq \drawmatrix[size=.1]M^{-1}
        \drawmatrix[height=.25, bbox width=1, width=.25]X
    $!
!\drawskip!
    \end{lstlisting}
\end{experiment}

\parsum{results}
For the selected range of $m$, the runtime of this experiment is below
\SI{200}{ms}, i.e., already more than 1 order of magnitude less than the
previous experiment.  Furthermore, for larger problems, the {\tt dpotrs} kernel
can make good use of the co-processor's many cores and reaches over
\SI{550}{Gflops/s}.  For an in depth study of performance optimizations for
GWAS we refer the reader to \cite{smp-gwas} and \cite{gwasalg}.


    \section{Conclusion}
    \label{sec:conclusion}
    We introduced the Experimental Linear Algebra Performance Studies framework
(\elaps), a set of tools and features to design, execute, measure and analyze
dense linear algebra performance experiments.   With its intuitive interface,
\elaps assists users in investigating performance behaviors and in making
informed decisions.  Throughout this paper, we applied \elaps to a wide range
of scenarios, including parameter optimization (\autoref{sec:sumrange}),
algorithm selection (\autoref{sec:ntrange}, \autoref{sec:tensors}), library
comparison (\autoref{sec:sylv}), and parallelism and library threading
(\autoref{sec:ex-omprange}, \autoref{sec:lus}).  We demonstrated the
framework's flexibility by linking it with seven kernel-libraries, and by
executing experiments on five different platforms, including the Xeon Phi
co-processor and two different batch-job systems.  In summary, \elaps covers
many aspects of shared-memory optimizations, a critical step towards achieving
large-scale performance.

Having established the foundations of a framework for rapid experimentation, we
foresee many opportunities for extensions. In particular, we envision 1)
coverage of a broader range of architectures, including GPUs and ARM-based
CPUs, 2) support for metrics related to data movement and energy consumption,
and 3) interfaces for distributed memory libraries such as {\sc ScaLAPACK}.


    \section*{Software}
    \elaps is open source (BSD license) and available on GitHub:
    \url{http://github.com/elmar-peise/ELAPS}

    \section*{Acknowledgments}
    Financial support from the Deutsche Forschungsgemeinschaft (DFG) through
    grant GSC 111 and from the Deutsche Telekom Stiftung, and access to JUQUEEN
    at J\"ulich Supercomputing Centre (JSC) are gratefully acknowledged.

    \bibliographystyle{abbrv}
    \bibliography{references}
\end{document}